\documentclass[twocolumn,aip,nofootinbib,jcp,reprint,amsmath,amssymb]{revtex4-2}

\usepackage{natbib}
\usepackage{amsmath}
\usepackage{graphicx}
\usepackage[hidelinks]{hyperref}
\usepackage{dcolumn}
\usepackage{bm}
\usepackage{epsfig,color,xspace,multirow,xr,bbold}
\usepackage[all]{xy}
\usepackage{setspace}
\usepackage{url}
\usepackage{soul}
\usepackage[colorinlistoftodos]{todonotes}
\usepackage{longtable}
\usepackage{tabularx}
\usepackage{adjustbox}
\usepackage{threeparttable} 
\usepackage{natbib,hyperref}
\usepackage{float}
\usepackage{placeins}
\usepackage[utf8]{inputenc}
\usepackage[T1]{fontenc}
\usepackage{pifont}

\newcommand{\cs}{$C_{\rm s}$\,}
\newcommand{\ctwov}{$C_{\rm 2v}$\,}
\newcommand{\cthreeh}{$C_{\rm 3h}$\,}

\newcommand{\dthreeh}{$D_{\rm 3h}$\,}

\usepackage[normalem]{ulem}
\usepackage{xcolor}
\newcommand{\RRef}[1]{Ref.~\onlinecite{#1}}

\newcommand\notsotiny{\@setfontsize\notsotiny\@vipt\@viipt}
\usepackage{soul} 
\newcommand{\orcid}[1]{\href{https://orcid.org/#1}{\includegraphics[width=8pt]{orcid.png}}}

\begin{document}

\title{
Insights into Symmetry and Substitution Patterns Governing Singlet–Triplet Energy Gap in the Chemical Space of Azaphenalenes
}
\date{\today}

\author{
Atreyee Majumdar, 
Raghunathan Ramakrishnan$^*$ 
}

\email{ramakrishnan@tifrh.res.in} 

\thanks{\\ $^*$ Corresponding author}  
\affiliation{Tata Institute of Fundamental Research, Hyderabad 500046, India}


\begin{abstract}
\noindent
Molecules that violate Hund’s rule by exhibiting an inverted singlet–triplet gap (STG), where the first excited singlet (S$_1$) lies below the triplet (T$_1$), are rare but hold great promise as efficient fifth-generation light emitters. Azaphenalenes (APs) represent one of the few known molecular classes capable of such inversion of the S$_1$/T$_1$ energy ordering, yet a systematic exploration of all unique APs is lacking. Here, we investigate 104 distinct APs and classify them based on their adherence to or deviation from Hund’s rule using S$_1$–T$_1$ gaps computed with the second-order coupled-cluster method employing the Laplace transform (L-CC2). To capture substitution-dependent pseudo–Jahn–Teller distortions that are inadequately described by MP2 and DFT methods, we employ focal-point extrapolation scheme to obtain near-CCSD(T)/cc-pVTZ-quality geometries. We find three APs to undergo $D_{\rm 3h} \rightarrow C_{\rm 3h}$ and ten to show $C_{\rm 2v}\rightarrow C_{\rm s}$ symmetry lowering, leading to a total of 117 configurations of 104 unique APs. Our study identifies top candidates with inverted STGs, revealing how substitution and symmetry-lowering modulate these gaps to uncover new stable AP cores that provide promising targets for designing molecular light-emitters.
\end{abstract}

\maketitle

\section{Introduction}\label{sec:introduction}

Organic light-emitting diodes (OLEDs) have attracted significant attention due to their remarkable emissive properties~\cite{uoyama2012highly}.
Successive generations of OLEDs have been developed, each addressing limitations of its predecessors~\cite{chen2018thermally}.
Third-generation OLEDs rely on thermally activated delayed fluorescence (TADF), in which reverse intersystem crossing (RISC) enables upconversion from the first triplet (T$_1$) to the first singlet excited state (S$_1$)~\cite{chen2018thermally,uoyama2012highly}.
Fourth-generation OLEDs, based on hyperfluorescence, combine TADF and fluorescence emitters but continue to suffer from efficiency losses due to slow RISC and long-lived triplet states, as well as stability concerns~\cite{li2022organic}.
However, the emerging fifth-generation emitters have the potential to deliver the most efficient light-emitting materials~\cite{aizawa2022delayed}. This new generation of emitters consists of molecules that show the delayed fluorescence from inverted singlet and triplet states (DFIST), leading to a down-conversion from T$_1$ to S$_1$, and the RISC becomes thermodynamically favourable\cite{aizawa2022delayed}. This unusual property of a negative S$_1$-T$_1$ gap (STG), goes against the Hund's rule which states that for a closed-shell system with S$_0$ ground state, T$_1$ should have a lower energy than S$_1$ \cite{toyota1986violation,borden1994violations}.

The theoretical origin of negative STGs arises from the interplay between exchange and correlation effects~\cite{jorner2024ultrafast,dreuw2023inverted}.
From an exchange-only picture based on self-consistent-field (SCF) orbitals, the S$_1$ and T$_1$ states are represented primarily by singly excited configuration state functions (CSFs), $^1\chi_{a \rightarrow r}$ and $^3\chi_{a \rightarrow r}$, where $a$ and $r$ denote the occupied and virtual molecular orbitals (MOs) obtained from the SCF calculation of the ground-state configuration, $^1\chi_0$. Typically, $a$ and $r$ correspond to the highest occupied and lowest unoccupied MOs (HOMO and LUMO), respectively. 
In the SCF formalism, the excitation energies of the S$_1$ and T$_1$ states
can be expressed in terms of singly excited CSFs as
\begin{align}
E_{{\rm S}_1}=E(^1\chi_{a \rightarrow r}) - E(^1\chi_0) &= \varepsilon_r - \varepsilon_a - J_{ar} + 2K_{ar}, \label{eq:STG1}\\
E_{{\rm T}_1}=E(^3\chi_{a \rightarrow r}) - E(^1\chi_0) &= \varepsilon_r - \varepsilon_a - J_{ar},
\label{eq:STG2}
\end{align}
where $J_{ar}$ and $K_{ar}$ are the Coulomb and exchange integrals, respectively. 
The STG, therefore, becomes
\begin{equation}
{\rm STG} = E(^1\chi_{a \rightarrow r}) - E(^3\chi_{a \rightarrow r}) = 2K_{ar}.
\end{equation}
As the spatial overlap between the densities of MOs $a$ and $r$ decreases, $K_{ar}$ diminishes, resulting in near-degeneracy of S$_1$ and T$_1$.\cite{leupin1980low,pu2025computational}

However, reducing $K_{ar}$ alone cannot invert the energy ordering.
In a more accurate description accounting for electron correlation effects, 
${\rm STG}$ contains additional terms that account for the way electrons dynamically adjust their spin densities to minimize Coulomb repulsion
\begin{equation}
{\rm STG} = 2K_{ar} + \Delta E_{\mathrm{corr(spin)}} ,
\end{equation}
where $\Delta E_{\mathrm{corr(spin)}}$ is the correlation-induced contribution associated with dynamic spin polarization (DSP)~\cite{jorner2024ultrafast,chanda2025benchmark,lashkaripour2025addressing}.
This term is intrinsically negative, $\Delta E_{\mathrm{corr(spin)}} < 0$, if DSP selectively stabilizes the singlet configuration through double excitations that lower its energy relative to the triplet.
This stabilization arises from singlet-specific coupling to higher-order excited configurations, which is absent for the triplet~\cite{bonacic1985charge,kollmar1978violation}.
The difference between the singlet and triplet correlation energies defines the spin-correlation component of the STG
\begin{equation}
\Delta E_{\mathrm{corr(spin)}} =
E_{\mathrm{corr}}(^1\chi_{a \rightarrow r}) -
E_{\mathrm{corr}}(^3\chi_{a \rightarrow r}).
\end{equation}
Note that, $E(^1\chi_{a\rightarrow r})$ and $E(^3\chi_{a\rightarrow r})$ denote the
total energies of the ${\rm S}_1$ and ${\rm T}_1$ excited states, respectively (see
Eqs.~\ref{eq:STG1} and \ref{eq:STG2}). Hence 
$E_{\mathrm{corr}}(^1\chi_{a\rightarrow r})$ and
$E_{\mathrm{corr}}(^3\chi_{a\rightarrow r})$ represent the corresponding
correlation contributions to these excited-state energies.
The sign of $\Delta E_{\mathrm{corr(spin)}}$ is therefore determined by the
relative stabilization of the singlet and triplet states due to electron
correlation.
In most systems, this difference is modest and does not alter the state ordering, but in molecules with very small exchange splittings, the enhanced singlet stabilization can render $\Delta E_{\mathrm{corr(spin)}}$ sufficiently negative to invert the STG. This criterion requires the opposite-spin correlation effects to be stronger than the same-spin correlation. Thus, while reducing orbital overlap lowers $K_{ar}$ and brings S$_1$ and T$_1$ closer in energy, DSP is the true physical origin of singlet-triplet inversion.

The emergence of negative STGs can be rationalized within the simplified Hubbard framework, where correlation effects are captured through an effective kinetic-exchange (superexchange) term~\cite{chanda2025benchmark}
\begin{equation}
{\rm STG} \approx 2K_{ar} - \frac{t^2}{\Delta E},
\end{equation}
where $t$ is the effective electron-hopping integral between sites (or orbitals) and $\Delta E$ is the energy penalty for double occupancy.
Here, the first term represents the direct (Hund-type) exchange that stabilizes the triplet, while the second 
term is analogous to the antiferromagnetic superexchange interaction term $(-t^2/U)$ in the Hubbard model~\cite{fazekas1999lecture}.
When the magnitude of this kinetic exchange term outweighs the exchange contribution,
\begin{equation}
\frac{t^2}{\Delta E} > 2K_{ar},
\end{equation}
the singlet lies below the triplet, resulting in a negative STG.

Leupin et al.~\cite{leupin1980low} first proposed the possibility of an inverted STG in cycl[3.3.3]azine, commonly known as cyclazine (1AP), a triangular polycyclic molecule. 
Hund’s rule violations manifested as negative STGs in 
both cyclazine and heptazine (7AP) 
were first predicted by computational studies~\cite{de2019inverted,ehrmaier2019singlet}. 
Notably, Ehrmaier et al. also provided experimental support for singlet–
triplet inversion in a heptazine derivative (tri-anisole heptazine) using
time-resolved photoluminescence measurements, where an unusually long
fluorescence lifetime persisted even in the presence of molecular oxygen, which
would normally quench triplet states \cite{wrigley2025singlet}.
A subsequent computational analysis highlighted potential methodological pitfalls affecting singlet–triplet inversions when lower-level approaches are used~\cite{dreuw2023inverted}. However, high-level 
electronic-structure calculations~\cite{loos2023heptazine,loos2025correction}, including those employing CC3, have unambiguously confirmed the existence of genuine inverted singlet–triplet gaps. More recently, the INVEST property was also confirmed for pentaazaphenalene (5AP)
through an experimental study based on anion photoelectron spectroscopy
\cite{wilson2024spectroscopic}.

An exhaustive search~\cite{majumdar2024resilience} of the structurally diverse small-molecule chemical space, bigQM7$\omega$, containing $\sim$13,000 molecules with CONF atoms~\cite{kayastha2022resolution}, revealed no violations of Hund’s rule. This suggests that achieving ${\rm STG}<0$ requires non-trivial molecular frameworks absent from conventional small-molecule chemical space~\cite{majumdar2024resilience}.


The present work undertakes a systematic study of all possible N-substitution patterns in APs, encompassing systems with nitrogen (N) atoms placed at electron-rich, electron-deficient, or both types of sites. 
To obtain accurate ground-state geometries efficiently, we employ a geometry scheme that achieves CCSD(T)-level accuracy through basis-set extrapolation of optimized internal coordinates.
Each AP is classified per point group, and symmetry-lowering cases are identified by tracing reductions from the highest symmetry permitted by composition and connectivity to their low-symmetry subgroups. STGs are then evaluated using the 

CC2 response theory with Laplace-transform-based treatment of the doubles contributions (L-CC2) 
to identify systems that violate Hund’s rule and to reveal substitution- and symmetry-dependent trends governing the gap. 
Extensive benchmarking showed L-CC2 to provide accurate excited-state energies for APs at a moderate computational cost~\cite{majumdar2025leveraging}. The remainder of this article presents qualitative analyses of the dataset, methodological details, and trends illustrating how substitution and symmetry-lowering modulate STGs across the chemical space of APs.

\begin{figure*}[!hptb]
\centering
\includegraphics[width=\linewidth]{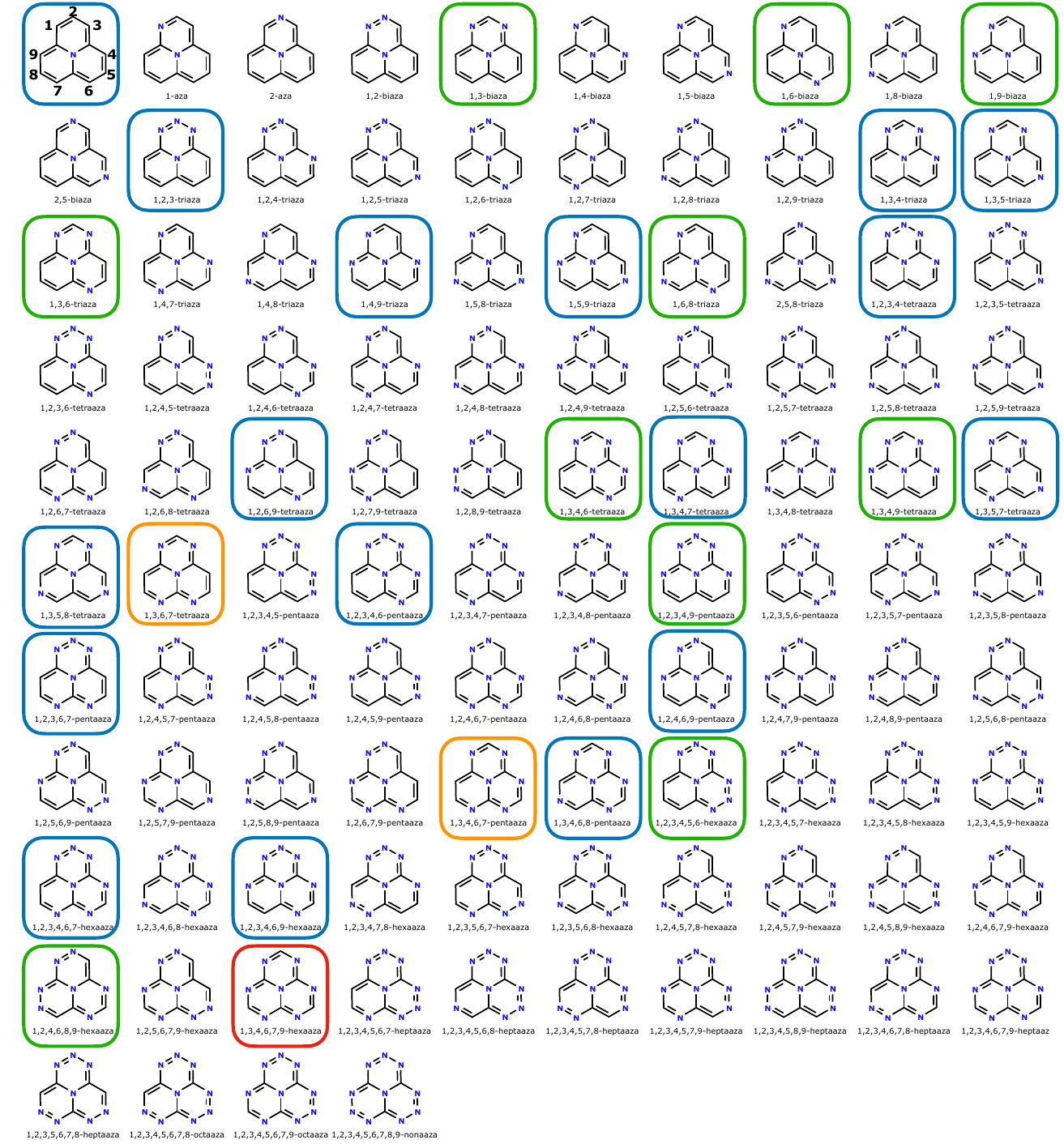}
\caption{
Representation of 104 azaphenalenes. 
The name of each molecule follows the convention {\it prefix}cycl[3.3.3]azine or {\it prefix}cyclazine, where {\it prefix} is stated below the structure. 
The first molecule (top, left) is cycl[3.3.3]azine or cyclazine without a prefix. 
Molecules with negative S$_1$-T$_1$ energy gaps predicted with L-CC2/aug-cc-pVDZ in the range 
$<-0.2$ eV, $\left(-0.2 , -0.15\right]$ eV,  $\left(-0.15 , -0.1\right]$ eV, 
and $(-0.1, 0.0]$ eV are shown in red, orange, green, and blue boxes. 
}
\label{fig_allsystems}
\end{figure*}

\section{Systems and Methods}

\subsection{Enumeration of azaphenalene chemical space\label{sec_enumeration}}
Unique molecular structures can be generated from Simplified Molecular Input Line Entry System (SMILES) strings~\cite{weininger1988smiles}, with redundant entries removed using tools such as OpenBabel~\cite{o2011open}. However, this procedure is not entirely foolproof and it fails to identify certain unique molecules, especially in the generation of chemical space datasets. 
A more reliable strategy for comprehensive coverage of chemical space is to mathematically enumerate all possible structures using combinatorial techniques and compare the resulting count with that obtained from enumeration by construction.

For molecules with a fixed molecular graph and atoms substituted in a combinatorial manner, the total number of unique APs can be determined using the P{\'o}lya enumeration technique~\cite{polya1937kombinatorische,polya2012combinatorial}. 
This method enables the systematic calculation of distinct isomers, expressed as a pattern inventory, and has been widely applied in the enumeration of chemical spaces~\cite{balasubramanian1985applications,balasubramanian1992combinatorics,faulon2005enumerating,balaban1991enumeration,chakraborty2019chemical}.

To apply the P{\'o}lya enumeration technique, the maximal molecular symmetry group ($\mathcal{G}$) of the graph, in which atoms are substituted combinatorially, must first be mapped onto an equivalent permutation group. For APs, $\mathcal{G}$ corresponds to the ${D}_3$ point group, which can be represented as the following set of permutations:
$E \equiv (1)(2)(3)(4)(5)(6)(7)(8)(9)$,
$C_3 \equiv (1,4,7)(2,5,8)(3,6,9)$,
$C_3^2 \equiv (1,7,4)(2,8,5)(3,9,6)$,
$C_2^\prime \equiv (1)(2,9)(3,8)(4,7)(5,6)$,
$C_2^{\prime\prime} \equiv (4)(3,5)(2,6)(1,7)(8,9)$,
$C_2^{\prime\prime\prime} \equiv (7)(2,3)(1,4)(5,9)(6,8)$.
The cycle structure of this group can be compactly expressed as
$\lbrace P_1^9, \, 2P_3^3, \, 3P_1^1 P_2^4 \rbrace$,
where $NP_n^m$ denotes $N$ permutations, each containing $m$ cycles of length $n$.

The pattern inventory ($\mathcal{Z}$) is defined as a function of the maximal symmetry molecular 
symmetry group, $\mathcal{G}$, as
follows
\begin{eqnarray}
\mathcal{Z} & = & \frac{1}{ |\mathcal{G}| }
\sum_{g \in \mathcal{G}} 
\left[ \Pi_{k} \mathcal{F}_k \right],
\label{eq:ci}
\end{eqnarray}
where $g$ spans all elements of the permutation group $\mathcal{G}$, and
$|\mathcal{G}|$ denotes the group order (which is 6 for the $D_3$ point group suitable to describe the triangular core of APs). The symbol
$\Pi_{k}$ denotes the product over each factor-cycle in the cycle representation of the element $g$, while
$\mathcal{F}_k$ is the figure counting series. 
The identity element is a product of 9 cycles of length 1, hence for the identity element,
$\Pi_{k}\mathcal{F}_k $ corresponds to $\left( {\rm C} + {\rm N} \right)^9$. 
Similarly, for the three-fold
rotations, $\Pi_{k}\mathcal{F}_k $ corresponds to $\left( {\rm C}^3 + {\rm N}^3 \right)^3$, while 
for the two-fold rotations, 
$\Pi_{k}\mathcal{F}_k $ corresponds to $\left( {\rm C} + {\rm N} \right)\left( {\rm C}^2 + {\rm N}^2 \right)^4$.
Introducing these expressions in Eq.~\ref{eq:ci} results in the  pattern inventory
\begin{eqnarray}
\mathcal{Z} & = & 
    \frac{1}{6} \left[ \left( {\rm C} + {\rm N} \right)^9  + 2 \left( {\rm C}^3 + {\rm N}^3 \right)^3 + 3 \left( {\rm C} + {\rm N} \right)\left( {\rm C}^2 + {\rm N}^2 \right)^4 \right]     \nonumber \\ 
    & = & 
    {\rm C}_9+2{\rm C}_8{\rm N}+8{\rm C}_7{\rm N}_2+17{\rm C}_6{\rm N}_3+24{\rm C}_5{\rm N}_4+\nonumber \\
& & 24{\rm C}_4{\rm N}_5+17{\rm C}_3{\rm N}_6+8{\rm C}_2{\rm N}_7+2{\rm C}{\rm N}_8+{\rm N}_9.
\label{eq:patterninventory}
\end{eqnarray}

The coefficient of each term in Eq.~\ref{eq:patterninventory} corresponds to the number of unique APs with a given composition, considering the nine substitution sites shown on the top, left of Figure~\ref{fig_allsystems}. For instance, the second term of Eq.~\ref{eq:patterninventory} ($2{\rm C}_8{\rm N}$) indicates that two unique APs contain a single N atom in the periphery of cyclazine. Similarly, the fourth term ($17{\rm C}_6{\rm N}_3$) shows that there are 17 distinct triazacyclazines (with three peripheral N atoms). Summing all coefficients in Eq.~\ref{eq:patterninventory} yields a total of 104 unique azaphenalenes. This count matches the set of unique SMILES strings filtered using OpenBabel, with the corresponding structures illustrated in Figure~\ref{fig_allsystems}.

All 104 APs shown in Figure~\ref{fig_allsystems} are named consistently using a standard numbering scheme, where the peripheral carbon atoms are labeled from 1 to 9. Each system is designated according to the positions occupied by N. Systems with N atoms exclusively at positions 1, 3, 4, 6, 7, or 9 are topologically charge-stabilized and are less susceptible to pseudo-Jahn--Teller distortions, as discussed in \RRef{majumdar2024influence}. In contrast, systems with N substitutions at positions 2, 5, or 8 are topologically charge-destabilized and undergo structural distortions to achieve a stable geometry~\cite{majumdar2024influence}. Mixed systems, containing N atoms in both stabilizing and destabilizing positions, are also possible; however, these have not previously been thoroughly examined in the context of their geometric stability and inverted STGs. In this work, we investigate all three categories of APs.

\subsection{Survey of APs}\label{sec:survey}

\begin{table*}[!htpb]
\centering
\caption{
List of azaphenalenes investigated in this work along with references to prior studies on these molecules. The N substitution sites are classified as electron-rich or electron-deficient positions on the cyclazine periphery. A few systems exhibit both stabilizing and destabilizing substitutions. Molecular structures
and naming conventions are given in Figure~\ref{fig_allsystems}.
}
\label{tab_exp_ref}
\addtolength{\tabcolsep}{1.2pt}
\begin{tabular}{l l l l c c}
\hline
S.No. & System & \multicolumn{2}{l}{Past works}& \multicolumn{2}{l}{Sites} \\
\cline{3-6}
              &                 & Experimental & Computational & $e^-$ rich & $e^-$ deficient \\
\hline
1&cyclazine$^a$         & \RRef{leupin1980low} & \RRef{loos2023heptazine,derradji2024searching,de2019inverted,tuvckova2022origin,bedogni2023shining}  &  & \\ 
2&1-aza & & \RRef{sabljic1978theoretical,kurihara1994cyclazines} & \ding{51}\\
3&2-aza & & \RRef{kurihara1994cyclazines,loos2023heptazine,alipour2022any,sancho2022violation,ghosh1995electronic} && \ding{51}\\
4&1,3-biaza & \RRef{flitsch1978cyclazines} & \RRef{kurihara1994cyclazines,alipour2022any,bedogni2023shining} & \ding{51}\\
5&1,4-biaza & \RRef{matsuda1987chemistry,ceder1977synthesis} & & \ding{51}\\
6&1,6-biaza & \RRef{matsuda1987chemistry,gotou1985studies} & & \ding{51}\\
7&1,8-biaza & & \RRef{ghosh1995electronic} & \ding{51} & \ding{51}\\
8&1,9-biaza  & \RRef{ceder1976synthesis} & \RRef{tuvckova2022origin,pollice2021organic,alipour2022any,sancho2022violation,loos2023heptazine} & \ding{51} \\
9&2,5-biaza & & \RRef{kurihara1994cyclazines,ghosh1995electronic,loos2023heptazine} & & \ding{51}\\
10&1,3,4-triaza & \RRef{ceder1977synthesis} & \RRef{tuvckova2022origin,pollice2021organic,sancho2022violation,alipour2022any,bedogni2023shining} & \ding{51}\\
11&1,3,6-triaza & \RRef{kanamori1992systematic,rossman1985synthesis} & \RRef{ghosh1995electronic} & \ding{51}\\ 
12&1,4,7-triaza & \RRef{leupin19861} & \RRef{bhattacharyya2021can,ricci2021singlet,ghosh2022origin,sancho2022violation} & \ding{51}\\
13&2,5,8-triaza & & \RRef{kurihara1994cyclazines,ghosh1995electronic,loos2023heptazine} & & \ding{51}\\
14&1,3,4,6-tetraaza & \RRef{wilson2024spectroscopic} & \RRef{loos2023heptazine,derradji2024searching,tuvckova2022origin} & \ding{51}\\ 
15&1,3,6,7-tetraaza & & \RRef{kurihara1994cyclazines,ghosh1995electronic} & \ding{51}\\
16&1,3,4,9-tetraaza & & \RRef{tuvckova2022origin,pollice2021organic,sancho2022violation,alipour2022any,loos2023heptazine} & \ding{51}\\
17&1,3,4,7-tetraaza && \RRef{ghosh1995electronic,kurihara1994cyclazines} & \ding{51}\\
18&1,3,4,6,7-pentaaza& \RRef{rossman1985synthesis,kanamori1992systematic} & \RRef{kurihara1994cyclazines,ghosh1995electronic,garner2024enhanced,yang2014structure} & \ding{51}\\
19&1,3,4,6,8-pentaaza & \RRef{kanamori1992systematic,rossman1985synthesis,shaw1977fused} & & \ding{51} & \ding{51} \\
20&1,2,4,6,7,9-hexaaza & \RRef{rossman1985synthesis} & & \ding{51} & \ding{51}\\
21&1,3,4,6,7,9-hexaaza & \RRef{rossman1985synthesis,ehrmaier2019singlet} & \RRef{de2019inverted,ehrmaier2019singlet,ricci2021singlet,ghosh2022origin,aizawa2022delayed,loos2023heptazine,alipour2022any,sun2024design,tuvckova2022origin,bedogni2023shining,derradji2024searching} & \ding{51}\\
22&1,2,3,4,6,7,9-heptaaza  & & \RRef{zheng2004tri} & \ding{51} & \ding{51}\\
23&1,2,3,4,5,6,7,9-octaaza & & \RRef{zheng2004tri,politzer2013tricyclic} & \ding{51} & \ding{51}\\
24&1,2,3,4,5,6,7,8,9-nonaaza & & \RRef{politzer2013tricyclic,zheng2004tri,wu2014improving} & \ding{51} & \ding{51}\\
\hline
\end{tabular}
\label{tab:survey}
\begin{tablenotes}
 \item 
 \footnotesize{$^a$ special case with no N atoms in the periphery.} 
 \end{tablenotes}
\end{table*}

In 1980, Leupin and Wirz reported an unexpectedly small STG of 0.04 eV for cyclazine, based on fluorescence and energy-transfer experiments, suggesting a possible inversion or near-degeneracy of S$_1$ and T$_1$~\cite{leupin1980low}. Decades later, De Silva revisited~\cite{de2019inverted} this system using modern computational methods such as ADC(2), predicting an STG  (using vertical $E_{{\rm S}_1}$ and $E_{{\rm T}_1}$ at the S$_0$ geometry) of $-0.16$ eV. In their study, TD-DFT with hybrid exchange–correlation functionals predicted $\text{STG} > 0$, revealing the limitations of conventional DFT in capturing INVEST. 
Ehrmaier et al.~\cite{ehrmaier2019singlet} investigated 7AP   (1,3,4,6,7,9-hexaazacyclazine; see Figure~\ref{fig_allsystems}), and confirmed inversion of the STG with the ADC(2) method predicting $-0.28$ eV, in agreement with other wavefunction-based methods. Subsequent computational studies~\cite{aizawa2022delayed,pios2021triangular,pollice2021organic,ghosh2022origin,sancho2022violation,loos2023heptazine,won2023inverted,bedogni2023shining,chen2024unveiling,majumdar2024influence,perez2025role,ricci2025enhancing} have 
identified other derivatives of APs exhibiting INVEST.

Notably, Aizawa et al.~\cite{aizawa2022delayed} computationally screened $\sim$35,000 molecules and identified the 1,3,4,6,7,9-hexaazacyclazine (7AP) derivative, HzTFEX$_2$,  as a promising DFIST candidate with a negative STG. They used temperature-dependent photoluminescence (PL) decay measurements and kinetic analysis of the transient PL data,  to fit the temperature dependence of the ISC and  RISC rate constants ($k_{\rm ISC}$ and $k_{\rm RISC}$, respectively) using the Arrhenius equation, finding a negative activation energy difference $\Delta E_{\text{ST}} = -11 \pm 2~\text{meV}$. This provided an indirect experimental evidence of a negative STG for a 7AP derivative. 

Another 7AP derivative, HAP-3MF, was synthesized and evaluated as an OLED emitter
by Adachi and co-workers~\cite{li2014thermally}, who demonstrated that it surpasses
the theoretical efficiency limit of conventional fluorescent OLEDs via TADF.
Further theoretical analysis showed that this molecule exhibits an inverted
STG, thereby violating Hund’s rule~\cite{sobolewski2021heptazine}.
Combined with the experimental time-resolved photoluminescence evidence
reported earlier for other heptazine derivatives~\cite{ehrmaier2019singlet},
these findings further establish 7AP-based systems as a prominent molecular
class exhibiting inverted STGs.
Similarly, Actis et al.~\cite{actis2023singlet} reported an inversion of approximately 0.2 eV in a polymeric 7AP, combining time-resolved electron paramagnetic resonance with steady-state optical spectroscopy.

Among the few direct experimental demonstrations of singlet–triplet inversion, pentaazaphenalene (1,3,4,6-tetraazacyclazine; 5AP) holds particular significance. Wilson et al.~\cite{wilson2024spectroscopic} provided the first direct spectroscopic evidence of Hund’s rule violation in the closed-shell organic molecule, 5AP. Using anion photoelectron spectroscopy, they measured $E_{{\rm S}_1}$ and $E_{{\rm T}_1}$ precisely, obtaining a negative STG of $-0.047(7)$ eV. This value agrees with ADC(2)-level 0$-$0 estimates of $-0.094$ eV and CC2-level value of $-0.085$ eV reported in ~\RRef{tuvckova2022origin}. Kusakabe et al.~\cite{kusakabe2024inverted} further demonstrated that a dialkylamine-substituted 5AP derivative, 5AP-N(C${12}$)$_2$, exhibits barrierless RISC and thermally activated ISC in transient PL experiments, consistent with a negative STG.

A particularly significant computational contribution was made by Loos et al.~\cite{loos2023heptazine}, who employed high-level wavefunction-based methods to investigate ten APs, including 
cyclazine (1AP), heptazine (7AP), 2-azacyclazine (2AP), 2,5-biazacyclazine (3AP), 2,5,8-triazacyclazine (4AP), 1,9-biazacyclazine, among other molecules, 
at high-symmetry geometries such as \dthreeh and \ctwov. 
Their work provided  theoretical best estimates (TBEs) of STGs based on CCSD(T)/cc-pVTZ-level geometries, revealing negative values for several systems and thereby establishing a consistent and physically grounded description of INVEST.

For large-scale screening of INVEST molecules, DFT within the TDDFT framework remains the most practical approach for estimating STGs. 
However, TDDFT based on standard DFT functionals such as, pure (i.e., semilocal), hybrid, and range-separated hybrid functionals, are unable to provide a quantitatively reliable description
of STGs.
Previous studies~\cite{ghosh2022origin,tuvckova2022origin,kondo2022singlet,majumdar2024resilience} have shown that incorporating explicit electron correlation, such as with second-order perturbation theory within double-hybrid DFT, is essential for achieving STG values consistent with correlated wavefunction-based methods. 

Chanda et al.~\cite{chanda2025benchmark} evaluated DFT and TDDFT methods for selected APs, finding that spin-scaled double-hybrid functionals successfully reproduced negative STGs. 
Their $\Delta$SCF calculations based on hybrid DFT functionals (e.g., B3LYP), where $E_{{\rm S}_1}$ was obtained using the maximum-overlap method and $E_{{\rm T}_1}$ from a separate triplet-state SCF calculation, also yielded negative STGs. This apparent agreement, however, was found to arise artifactually from spin contamination in the S$_1$ state than in the T$_1$ state. Interestingly, while such spin contamination reflects an unphysical mixing of multiple CSFs within a single-determinant framework, it nonetheless indicates the inherent configuration mixing character of the S$_1$ state. This is consistent with the discussion in Section~\ref{sec:introduction}, where selective stabilization of S$_1$ relative to T$_1$ was indicated as a prerequisite for realizing a genuinely negative STG.

In addition to electron correlation, accurate treatment of molecular geometry and pseudo-Jahn--Teller (pJT) effects is critical for describing STG inversion. Loos et al.~\cite{loos2023heptazine} also reported certain molecules to exhibit negative STGs only at high-symmetry geometries; upon relaxation, symmetry-lowering restored positive gaps. For instance, cycloborane displayed an apparent negative STG at its \dthreeh saddle-point geometry, which became positive at the true minimum with \cthreeh symmetry. 

Similar behavior has been reported for B,N-substituted polycyclic aromatic
hydrocarbons~\cite{majumdar2025unlocking}, where high-symmetry transition-state
structures exhibit negative STGs that become positive upon relaxation to the
true minima. An early work also showed that propalene, pentalene, and heptalene in their
high-symmetry geometries possess a lowest excited singlet state lying below the
corresponding triplet~\cite{koseki1985violation}.
More recently, high-level calculations have demonstrated~\cite{loos2025quest} that this qualitative behavior, i.e., negative STGs at high-symmetry structures that revert to
positive values upon symmetry lowering, persists for non-alternant hydrocarbons
and bicyclic molecules, in agreement with earlier investigations~\cite{terence2023symmetry}.

Detailed analysis of six triangular candidates~\cite{majumdar2024influence} confirmed that only two retained both their equilibrium geometry and negative STGs, while the rest underwent pJT-driven distortions. The degree of distortion was found to depend strongly on the level of theory. 
Specifically, MP2 failed to capture pJT coupling, CCSD enhanced it, and CCSD(T) provided a balanced picture. 
Among DFT functionals, B3LYP and $\omega$B97XD captured the interaction partially but lacked quantitative accuracy relative to CCSD(T).
Collectively, these studies demonstrate that realizing a genuinely negative STG requires: 
(i) a correlated S$_1$ state exhibiting enhanced configuration mixing that selectively stabilizes it relative to T$_1$, and (ii) structural robustness against pJT distortion. Recent benchmarks using 
local CC2 and ADC(2) methods with Laplace-transform-based formulations
further confirm that these correlated methods offer an efficient and accurate framework for screening medium-sized molecules such as azaphenalenes and boraphenalenes~\cite{majumdar2025leveraging}.

\begin{figure}[!hptb]
\centering
\includegraphics[width=\linewidth]{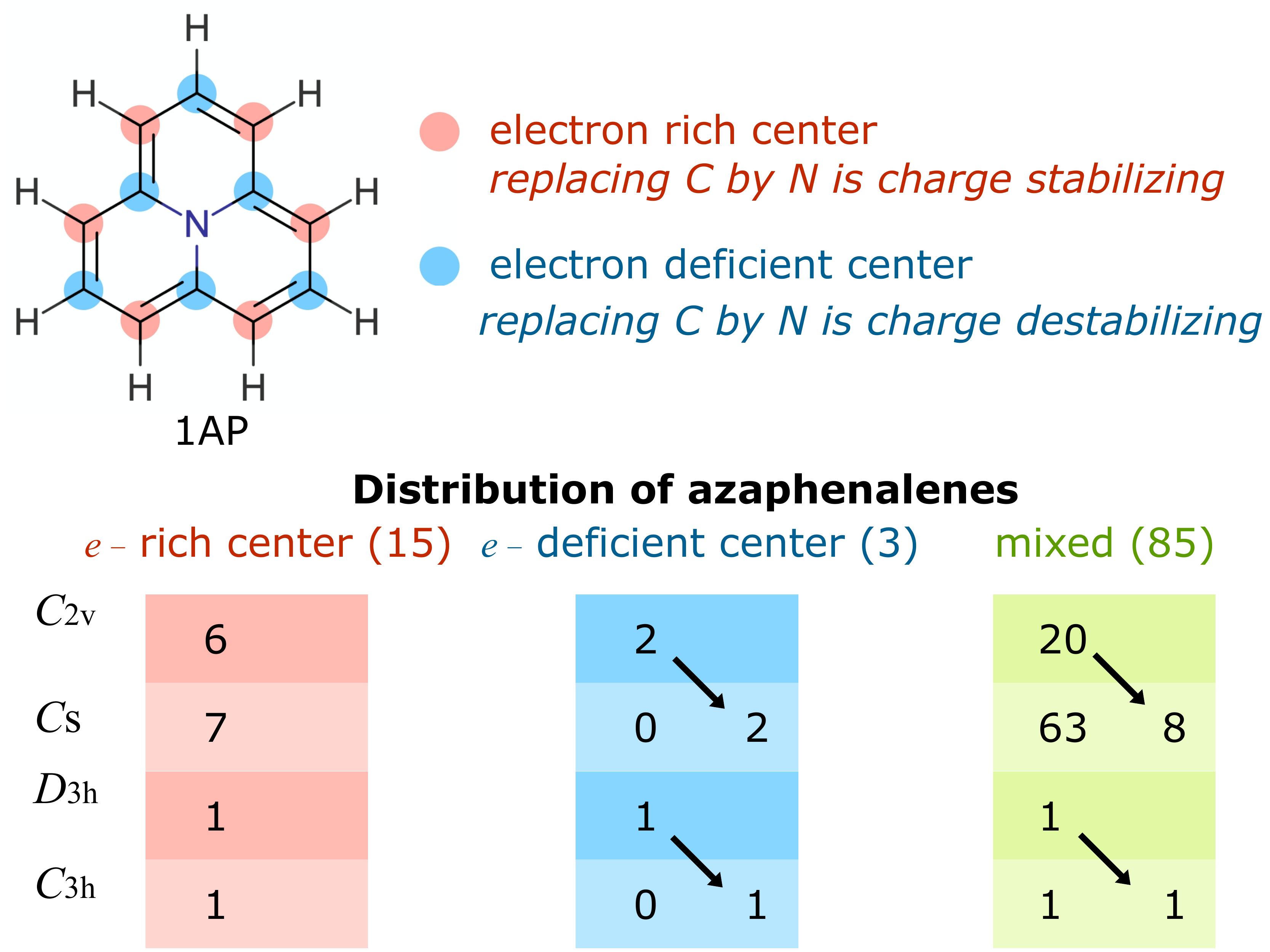}
\caption{
HOMO and LUMO density distributions along the periphery of cyclazine, illustrating the classification of substitution sites as electron-rich (HOMO-dominated) or electron-deficient (LUMO-dominated). The complete set of APs, derived from these substitution patterns are categorized according to the maximum symmetry point group allowed by their composition; arrows indicate cases of symmetry lowering. Cyclazine, a special case undergoing \dthreeh$\rightarrow$\cthreeh distortion, is not included in any of these classes, as it contains no N atoms along the periphery. 
}
\label{fig:sites}
\end{figure}

Table~\ref{tab:survey} summarizes the APs that have been investigated previously, along with corresponding references. The 24 systems listed represent a subset of the 104 APs considered in this work (see Figure~\ref{fig_allsystems}). The observed substitution patterns can be rationalized using the concept of topological charge stabilization~\cite{gimarc1983topological}, which posits that the placement of electronegative atoms (such as N) at electron-rich sites of the unsubstituted parent system (regions of high HOMO density) stabilizes the molecular framework. In contrast, such a substitution at electron-deficient (LUMO) sites tends to destabilize it. See Figure~\ref{fig:sites} for the distribution of the electron-rich/deficient sites in the peripheral framework of AP. The details of symmetry lowering are 
discussed in Section~\ref{sec:pJT}.

To quantify the effect of topological charge stabilization through an independent descriptor, nucleus-independent chemical shift (NICS) analysis was performed in ~\RRef{majumdar2024influence} for cyclazine (1AP), two APs with N substitution at electron-rich sites (1,3,4,6-tetraazacyclazine; 5AP and 1,3,4,6,7,9-hexaazacyclazine; 7AP), and three APs with N at electron-deficient sites: 2-azacyclazine (2AP), 2,5-biazacyclazine (3AP), and 2,5,8-triazacyclazine (4AP). The NICS(1)${\rm iso}$ values of 5AP and 7AP were found to have very small magnitudes, indicating that these systems are essentially non-aromatic with localized charge density. In contrast, the other four systems, at their symmetric geometries, exhibited large positive NICS(1)${\rm iso}$ values (evaluated 1~\AA{} above the ring centroid), characteristic of antiaromatic behavior~\cite{karadakov2008ground}. The relative trend followed 1AP~$<$~2AP~$<$~3AP~$<$~4AP.

This analysis further suggests that 1AP is also susceptible to structural distortion despite lacking N atoms at topologically destabilizing (electron-deficient) sites. This behavior can be rationalized by comparison with its parent hydrocarbon, the phenalenyl radical, from which 1AP is derived by N substitution at the central C atom. The LUMO of the phenalenyl radical belongs to the $a_2''$ irreducible representation of the \dthreeh point group. The $z$-basis function of \dthreeh, corresponding to the central site, also transforms as $a_2''$, making the contribution of the central atom to the LUMO symmetry-allowed. Consequently, N substitution at this position corresponds to an electron-deficient (LUMO) site, whereas the singly occupied MO (SOMO, $a_1''$) of the radical is localized only along the periphery. 
Upon distortion to a \cthreeh geometry, both the HOMO and LUMO of 1AP transform as $a''$, enabling a stabilizing interaction with the HOMO that outweighs the destabilizing mixing with the LUMO~\cite{majumdar2024influence}.

Overall, of the 24 previously studied APs listed in Table~\ref{tab:survey}, including cyclazine, fourteen feature peripheral N atoms at electron-rich positions, three at electron-deficient sites, and six exhibit mixed substitution patterns. Within the broader class of APs, those with N substitution at electron-rich positions have been studied far more extensively than those substituted at electron-deficient sites. Notably, 1,4,9-triazacyclazine, which is part of the chemical space of APs, has not been investigated to date, either experimentally or theoretically, despite featuring N substitution at electron-rich sites that, according to Gimarc’s topological charge stabilization principle~\cite{gimarc1983topological}, should confer enhanced stability. Among the 85 APs with N at both electron-rich and electron-deficient sites, only six have been studied before (see Table~\ref{tab:survey}). By systematically examining the remaining 79 unexplored molecules, our exhaustive enumeration of APs substantially expands the pool of candidate molecules for identifying potential INVEST systems.

\subsection{Basis set extrapolation scheme for geometries\label{sec:geom_scheme}}

Complete basis set (CBS) extrapolation is typically done for total electronic energy \cite{halkier1998basis}. 
A similar scheme has also been performed for energy gradients, dubbed ``gradient scheme''\cite{heckert2005molecular}. 
Basis set extrapolations have also been performed for geometrical parameters (i.e., ``geometry scheme'') such as bond lengths and bond angles, using carefully selected internal coordinates as shown in \RRef{puzzarini2009extrapolation}. 

For the minimum-energy geometries of APs, our desired accuracy corresponds to the CCSD(T)/cc-pVTZ level, as more approximate methods fail to capture pJT effects adequately. However, full geometry optimizations at the CCSD(T)/cc-pVTZ level are computationally prohibitive. Therefore, for 104 similar systems, we apply the geometry scheme for basis set extrapolation of geometric parameters as follows.
\begin{eqnarray}
    \textbf{X}^{\rm CCSD(T),\,\Delta_{DT}}_{\rm VTZ} & \approx &
    \textbf{X}^{\rm CCSD(T)}_{\rm VDZ} + \left[\textbf{X}^{\rm CCSD}_{\rm VTZ} - \textbf{X}^{\rm CCSD}_{\rm VDZ} \right],
    \label{eq:extraopolation} 
\end{eqnarray}
where the superscript $\Delta_{\rm DT}$ 
denotes estimating CCSD(T)/cc-pVTZ
level internal coordinates, $X=r$ (distances) and $X=\theta$ (angles) defining the 
Z-matrix using CCSD values (with cc-pVDZ and cc-pVTZ basis sets), and 
CCSD(T) values (with cc-pVDZ basis set).

The Z-matrix representation ensures translational and rotational invariance of the structural parameters across methods, allowing the extrapolation to act only on molecular internal coordinates that are purely vibrational degrees of freedom.
Furthermore, this representation preserves the planarity of all AP structures examined in this study. We validate the accuracy of the geometries obtained through this geometry scheme for ten benchmark systems previously studied in \RRef{majumdar2024influence}.
Specifically, the STGs computed on these extrapolated geometries show excellent agreement with those obtained using geometries optimized at the full CCSD(T)/cc-pVTZ level (i.e., without extrapolations). For comparison, the accuracy of geometries optimized using DFT methods (B3LYP and $\omega$B97XD) and MP2 with the cc-pVTZ basis set, as reported in \RRef{majumdar2024influence},
was also evaluated, revealing that while these methods capture general trends, they deviate in describing pJT distortions and other fine structural details relative to the CCSD(T)/cc-pVTZ reference. These aspects are discussed in detail in Section~\ref{sec:accuracygeom}.

\subsection{CCSD(T) Diagnostics to probe symmetry lowering\label{sec:pjt}}

It is essential to compute geometries at the CCSD(T) level and to verify that the structures correspond to true minima rather than transition states. Frequency calculations are typically used for this purpose, and if a transition state is identified, a normal-mode analysis is required to locate the corresponding minimum energy configuration. However, vibrational frequency calculations at the CCSD(T) level are computationally prohibitive for medium-sized systems (with 10--30 atoms) such as APs. 
Further, high-symmetry structures that are saddle points on the potential energy surface (PES) cannot serve as effective starting points for locating lower-symmetry minima, since atomic forces vanish at the saddle points.
To address these aspects, \RRef{majumdar2024influence} introduced a diagnostic approach to assess structural stability imparted by post-MP2-level correlation correction without requiring frequency calculations and full-dimensional normal-mode analysis.

Accordingly, constrained optimizations were first carried out on high-symmetry structures (\dthreeh and \ctwov), with internal coordinates relaxed in Z-matrix representation at the MP2/cc-pVDZ level. Two interatomic distances that contribute to the soft-vibrational mode were scanned, with $r_1 \in [1.30, 1.50]$\AA{} and $r_2 \in [1.30, r_1]$\AA{}, in increments of 0.01 \AA, to construct a two-dimensional PES at the same level. 
Higher-level single-point CCSD(T)/cc-pVDZ energies were then computed on the same PES to probe whether post-MP2-level electron correlation corrections suggest possible symmetry-lowering. 
When this analysis revealed low-symmetry structures that are more stable than the high-symmetry ones, the geometry corresponding to the identified pair of $(r_1,r_2)$ values was used as the starting point for CCSD(T)-level 
full geometry optimization (i.e., without any constraints) to locate the true minimum.

\subsection{Computational details\label{sec:comp_details}}
Full geometry optimizations were performed for all 104 APs shown in Figure~\ref{fig_allsystems}, under point-group symmetry constraints, with internal coordinates defined through Z-matrix representations. For each molecule, the internal coordinates were carefully selected to preserve the intended point-group symmetry without introducing redundancies. Geometry optimizations were carried out at both the highest allowed point group symmetry and its largest subgroups, namely ${C}_{3h} \subset {D}_{3h}$ and ${C}_{s} \subset {C}_{2v}$.
For systems prone to symmetry lowering, initial geometries were generated through pJT diagnostic analysis discussed in Section~\ref{sec:pJT}. The initial structures of all 104 APs were first obtained using the universal force field (UFF) in OpenBabel. 
Independent geometry optimizations were performed at the CCSD/cc-pVDZ, CCSD/cc-pVTZ, and CCSD(T)/cc-pVDZ levels. The resulting geometries were then combined using the geometry scheme (Eq.~\ref{eq:extraopolation}) to obtain extrapolated CCSD(T)/cc-pVTZ-quality coordinates for all systems. Overall, of the 104 APs, 13 were found to undergo pJT-driven distortions, leading to a total of 117 distinct configurations analyzed in this work.

For systems exhibiting pJT distortions, complete basis set (CBS) extrapolations were performed to estimate the automerization barrier, defined as the energy difference between the high-symmetry saddle point and the corresponding low-symmetry minima. The extrapolation was carried out by combining CCSD(T) energies obtained with the cc-pVTZ and cc-pVQZ basis sets, where the Hartree--Fock (HF) energy from the cc-pVQZ basis was taken as the reference and the correlation energy extrapolated according to
\begin{equation}
E_{\mathrm{corr}}^n = E_{\mathrm{corr}}^{\mathrm{CBS}} + \alpha n^{-3},
\end{equation}
where $n$ is the cardinal number of the basis set.

It is well established that CC3 yields very accurate STGs, but is computationally prohibitive for large systems~\cite{loos2023heptazine}. As shown in \RRef{loos2023heptazine,loos2025correction}, the CC2/aug-cc-pVTZ method provides the best balance between accuracy and cost, with a mean absolute error (MAE) and standard deviation of the error (SDE) of $0.013|0.011$~eV relative to TBE reference values. Benchmarking against twelve triangulene molecules \RRef{majumdar2025leveraging} demonstrated that 
the density-fitted local CC2 response method based on the Laplace transform
(L-CC2)~\cite{kats2006local,freundorfer2010local} with the aug-cc-pVDZ basis set, offers accuracy (MAE$|$SDE = $0.016|0.013$~eV) comparable to CC2 with substantial computational savings. 
L-CC2 offers significant speedups over canonical CC2 while retaining similar accuracy.
Accordingly, the S$_1$ and T$_1$ excitation energies for all APs were computed using L-CC2 with the aug-cc-pVDZ basis set.

All CCSD and CCSD(T) geometry optimizations, as well as L-CC2 and CCSD(T)/CBS energy calculations, were performed using Molpro (version~2015.1)~\cite{werner2015molpro}. 
Frontier MO and harmonic frequency analyses were performed at the $\omega$B97XD/cc-pVTZ level using Gaussian (version~16~C.01)~\cite{frisch2016gaussianshort}.

\section{Results and Discussions}
\label{sec:results}

\subsection{Accuracies of geometries}\label{sec:accuracygeom}
As a first step in our investigation, we evaluated the accuracy of the geometry scheme used for basis-set extrapolation of minimum-energy structures to the CCSD(T)/cc-pVTZ level. For this purpose, we selected the previously studied~\cite{majumdar2024influence} APs, 1,3,4,6-tetraazacyclazine (5AP) and 1,3,4,6,7,9-hexaazacyclazine (7AP) in their symmetric geometries, along with cyclazine (1AP), 2-azacyclazine (2AP), 2,5-biazacyclazine (3AP), and 2,5,8- triazacyclazine (4AP) in both their high-symmetry (saddle-point) and symmetry-lowered (minimum-energy) configurations. 
For each of the ten systems, geometry optimizations were performed at the CCSD/cc-pVDZ, CCSD/cc-pVTZ, and CCSD(T)/cc-pVDZ levels using Z-matrix representations (see Data Availability to access the optimized coordinates). After obtaining these three sets of geometries, Eq.~\ref{eq:extraopolation} was applied to generate extrapolated geometries of CCSD(T)/cc-pVTZ quality. STGs were then computed at the L-CC2/aug-cc-pVDZ level for all systems using these extrapolated geometries.

\begin{table*}[!htpb]
\centering
\caption{
Excited-state energies and error metrics for benchmark systems obtained using various methods for determining minimum-energy geometries. L-CC2/aug-cc-pVDZ excitation energies (S$_1$ and T$_1$ relative to S$_0$) and singlet-triplet gaps (S$_1$–T$_1$) are reported in eV. Molecular names are followed by their point group symmetries in parentheses. 
Error metrics (in eV) are computed relative to L-CC2/aug-cc-pVDZ energies evaluated at CCSD(T)/cc-pVTZ reference geometries ($\textbf{X}^{\rm CCSD(T)}_{\rm VTZ}$ ) from \RRef{majumdar2024influence}: MSE (mean signed error), MAE (mean absolute error), and SDE (standard deviation of error). MAE values are obtained by subtracting the reference values from each method’s results.
 }
 \label{tab_table}
\scriptsize
\addtolength{\tabcolsep}{1.2pt}
\begin{tabular}[t]{l rrrl rrrl rrrl rrrl rrr}
\hline
\multicolumn{1}{l}{System}   &
\multicolumn{4}{l}{$\textbf{X}^{\rm CCSD}_{\rm VDZ}$} & 
\multicolumn{4}{l}{$\textbf{X}^{\rm CCSD}_{\rm VTZ}$} & 
\multicolumn{4}{l}{$\textbf{X}^{\rm CCSD(T)}_{\rm VDZ}$} & 
\multicolumn{4}{l}{$ \textbf{X}^{\rm CCSD(T),\,\Delta_{DT}}_{\rm VTZ}$} & 
\multicolumn{3}{l}{ $\textbf{X}^{\rm CCSD(T)}_{\rm VTZ}$ } \\
\cline{2-4} \cline{6-8}  \cline{10-12}   \cline{14-16}   \cline{18-20} 
\multicolumn{1}{l}{} &
\multicolumn{1}{l}{S$_1$} &
\multicolumn{1}{l}{T$_1$} &
\multicolumn{1}{l}{S$_1$-T$_1$} &
\multicolumn{1}{l}{} &
\multicolumn{1}{l}{S$_1$} &
\multicolumn{1}{l}{T$_1$} &
\multicolumn{1}{l}{S$_1$-T$_1$} &
\multicolumn{1}{l}{} &
\multicolumn{1}{l}{S$_1$} &
\multicolumn{1}{l}{T$_1$} &
\multicolumn{1}{l}{S$_1$-T$_1$} &
\multicolumn{1}{l}{} &
\multicolumn{1}{l}{S$_1$} &
\multicolumn{1}{l}{T$_1$} &
\multicolumn{1}{l}{S$_1$-T$_1$} &
\multicolumn{1}{l}{} &
\multicolumn{1}{l}{S$_1$} &
\multicolumn{1}{l}{T$_1$} &
\multicolumn{1}{l}{S$_1$-T$_1$} \\
\hline 

cyclazine (\dthreeh)  &  $1.008$  &  $1.149$  &  $-0.141$  &&  $1.053$  &  $1.184$  &  $-0.131$  &&  $0.990$   &  $1.132$  &  $-0.142$  && $1.033$  &  $1.167$  &  $-0.134$  &&  $1.031$  &  $1.165$  &  $-0.134$  \\
cyclazine (\cthreeh)  &  $1.350$   &  $1.298$  &  $0.052$   &&  $1.384$  &  $1.331$  &  $0.053$   &&  $1.143$  &  $1.193$  &  $-0.050$   &&  $1.181$  &  $1.228$  &  $-0.047$  &&  $1.162$  &  $1.217$  &  $-0.055$  \\
2-aza (\ctwov)  &  $0.856$  &  $0.946$  &  $-0.09$   &&  $0.916$  &  $1.001$  &  $-0.085$  &&  $0.832$  &  $0.926$  &  $-0.094$  &&  $0.889$  &  $0.978$  &  $-0.089$  &&  $0.887$  &  $0.975$  &  $-0.088$  \\
2-aza (\cs)   &  $1.388$  &  $1.203$  &  $0.185$   &&  $1.436$  &  $1.249$  &  $0.187$   &&  $1.180$   &  $1.087$  &  $0.093$   &&  $1.224$  &  $1.129$  &  $0.095$   &&  $1.211$  &  $1.120$   &  $0.091$   \\
2,5-biaza (\ctwov)  &  $0.725$  &  $0.799$  &  $-0.074$  &&  $0.771$  &  $0.839$  &  $-0.068$  &&  $0.699$  &  $0.774$  &  $-0.075$  &&  $0.737$  &  $0.808$  &  $-0.071$  &&  $0.734$  &  $0.806$  &  $-0.072$  \\
2,5-biaza (\cs)   &  $1.425$  &  $1.140$   &  $0.285$   &&  $1.481$  &  $1.193$  &  $0.288$   &&  $1.237$  &  $1.016$  &  $0.221$   &&  $1.289$  &  $1.065$  &  $0.224$   &&  $1.284$  &  $1.063$  &  $0.221$   \\
2,5,8-triaza (\dthreeh)  &  $0.584$  &  $0.643$  &  $-0.059$  &&  $0.620$   &  $0.675$  &  $-0.055$  &&  $0.556$  &  $0.622$  &  $-0.066$  &&  $0.593$  &  $0.651$  &  $-0.058$  &&  $0.590$   &  $0.646$  &  $-0.056$  \\
2,5,8-triaza (\cthreeh) &  $1.493$  &  $1.116$  &  $0.377$   &&  $1.555$  &  $1.179$  &  $0.376$   &&  $1.313$  &  $0.978$  &  $0.335$   &&  $1.372$  &  $1.035$  &  $0.337$   &&  $1.366$  &  $1.03$   &  $0.336$   \\
1,3,4,6-tetraaza (\ctwov)  &  $2.165$  &  $2.307$  &  $-0.142$  &&  $2.223$  &  $2.350$   &  $-0.127$  &&  $2.132$  &  $2.283$  &  $-0.151$  &&  $2.190$   &  $2.326$  &  $-0.136$  &&  $2.188$  &  $2.324$  &  $-0.136$  \\
1,3,4,6,7,9-hexaaza (\dthreeh)  &  $2.701$  &  $2.943$  &  $-0.242$  &&  $2.758$  &  $2.981$  &  $-0.223$  &&  $2.665$  &  $2.927$  &  $-0.262$  &&  $2.723$  &  $2.952$  &  $-0.229$  &&  $2.719$  &  $2.949$  &  $-0.230$   \\
MSE & $0.052$ & $0.025$ & $0.027$ && $0.102$ & $0.069$ & $0.034$ && $-0.042$ & $-0.036$ & $-0.007$ && $0.006$ & $0.004$ & $0.002$ \\
MAE & $0.074$ & $0.040$ & $0.034$ && $0.102$ & $0.069$ & $0.034$ && $0.042$ & $0.036$ & $0.008$ && $0.006$ & $0.004$ & $0.002$ \\
SDE & $0.088$ & $0.047$ & $0.043$ && $0.087$ & $0.051$ & $0.040$ && $0.012$ & $0.010$ & $0.010$ && $0.005$ & $0.003$ & $0.003$ \\
\hline
\end{tabular}
\label{tab:geom_benchmark}
\end{table*}%

Table~\ref{tab:geom_benchmark} summarizes the L-CC2/aug-cc-pVDZ excitation energies of the S$_1$ and T$_1$ states, along with the corresponding STGs, for the benchmark systems optimized at the CCSD/cc-pVDZ, CCSD/cc-pVTZ, and CCSD(T)/cc-pVDZ levels, as well as for the extrapolated CCSD(T)/cc-pVTZ geometries obtained using the geometry scheme described in Section~\ref{sec:geom_scheme}. For comparison, the CCSD(T)/cc-pVTZ geometries of the same systems were collected from ~\RRef{majumdar2024influence}, and their corresponding L-CC2/aug-cc-pVDZ excitation energies and STGs were evaluated in the present work.
The accuracies of all four sets of geometries were assessed relative to these reference data, with deviations quantified using the mean signed error (MSE), MAE, and SDE. The results presented in Table~\ref{tab:geom_benchmark} show that the L-CC2 values of S$_1$, T$_1$, and STG computed using the extrapolated geometries, $\textbf{X}^{\rm CCSD(T),\,\Delta_{DT}}_{\rm VTZ}$, are in excellent agreement with those obtained from the fully optimized $\textbf{X}^{\rm CCSD(T)}_{\rm VTZ}$ geometries. The magnitudes of the errors clearly demonstrate that this geometry extrapolation scheme yields systematically smaller deviations across all error metrics, confirming its reliability for large-scale data generation.

\begin{figure}[!hptb]
\centering
\includegraphics[width=\linewidth]{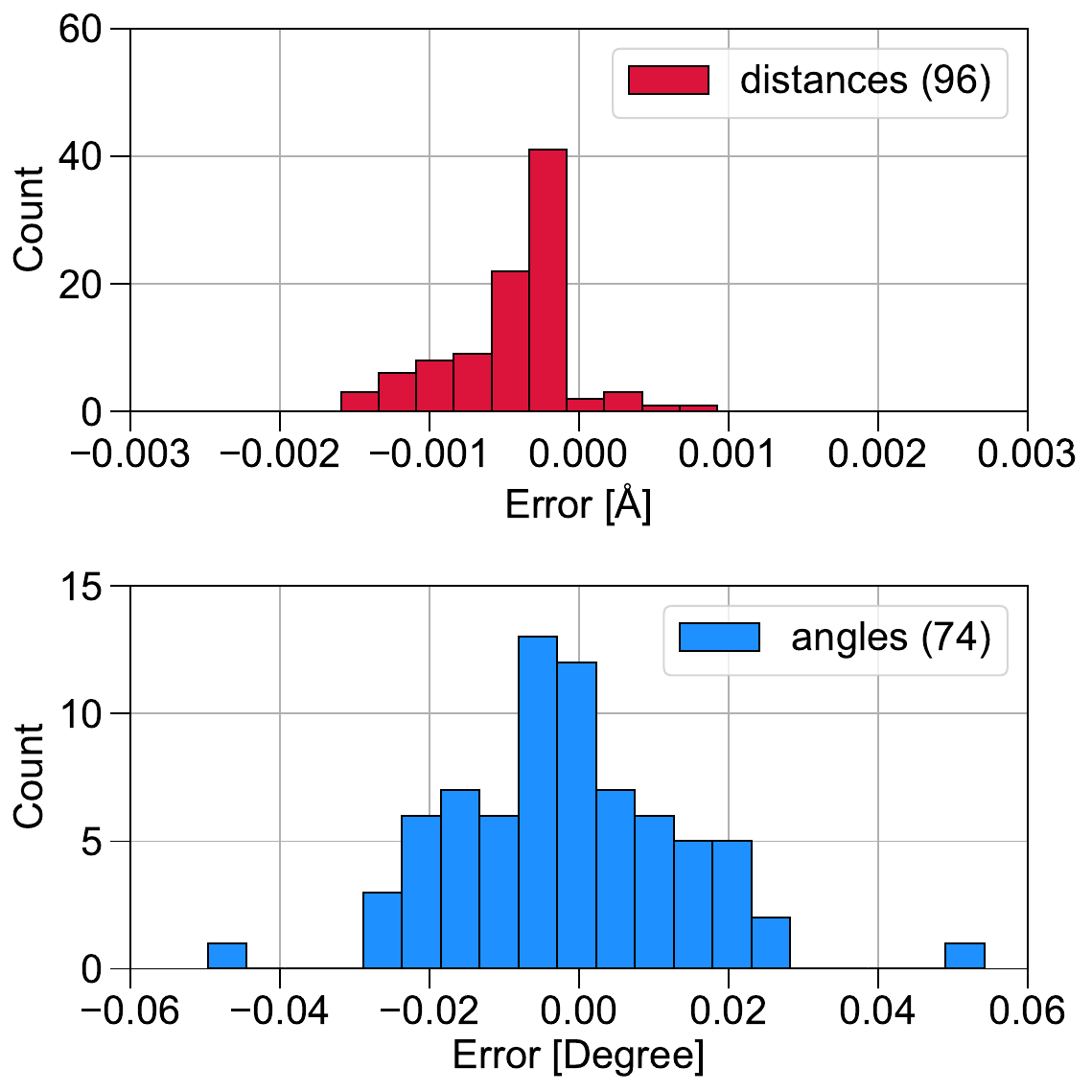}
\caption{
Histograms of errors in bond lengths and bond angles of the Z-matrix parameters for geometries optimized using
$ \textbf{X}^{\rm CCSD(T),\,\Delta_{DT}}_{\rm VTZ}$  relative to the reference
$ \textbf{X}^{\rm CCSD(T)}_{\rm VTZ}$  for the benchmark systems listed in Table~\ref{tab:geom_benchmark}.
}
\label{fig_dist_angle}
\end{figure}

From Table~\ref{tab:geom_benchmark}, it is evident that the extrapolated geometries, $\textbf{X}^{\rm CCSD(T),\,\Delta_{DT}}_{\rm VTZ}$, yield the most accurate L-CC2 STGs, with MSE, MAE, and SDE values of $0.002$, $0.002$, and $0.003$~eV, respectively. This clearly demonstrates that post-CCSD corrections to geometries are essential for obtaining reliable excitation energies.
Figure~\ref{fig_dist_angle} presents histograms of the deviations in bond distances and bond angles of the extrapolated geometries (Eq.~\ref{eq:extraopolation}) relative to the reference CCSD(T)/cc-pVTZ structures. The majority of molecules exhibit tiny deviations, with bond-length errors confined to the range of $-0.0015$ to $+0.001$~\AA{} and bond-angle errors concentrated between $-0.03^{\circ}$ and $+0.03^{\circ}$. 

The only notable exception is cyclazine in its \cthreeh geometry, which shows
a slightly larger angular deviation of about $0.05^{\circ}$.
At the CCSD(T)/cc-pVTZ level, cyclazine is characterized by a very shallow
PES with respect to symmetry-lowering distortions.
Since CCSD is known to over-stabilize the low-symmetry structure for this system,
the basis-set correction $\Delta_{\rm DT}^{\rm VTZ}$ derived from CCSD introduces a
somewhat larger residual error for cyclazine compared to the other molecules.
Nevertheless, this deviation remains small in absolute terms and does not alter
the qualitative conclusions. In consistent with this trend,
Table~\ref{tab:geom_benchmark} shows that the difference in STG
between the $\textbf{X}^{\rm CCSD(T),\,\Delta_{DT}}_{\rm VTZ}$ and
$\textbf{X}^{\rm CCSD(T)}_{\rm VTZ}$ geometries is largest for cyclazine
(in \cthreeh symmetry), but still amounts to only $0.008$ eV.

The primary contribution to the accuracy of $\textbf{X}^{\rm CCSD(T),\,\Delta_{DT}}_{\rm VTZ}$ arises from the baseline CCSD(T)/cc-pVDZ geometries, which already produce low error metrics (MSE, MAE, and SDE of $-0.007$, $+0.008$, and $+0.010$~eV, respectively). Although the CCSD/cc-pVDZ and CCSD/cc-pVTZ geometries exhibit slightly larger deviations, the basis-set correction incorporated through the extrapolation scheme effectively brings $\textbf{X}^{\rm CCSD(T),\,\Delta{DT}}_{\rm VTZ}$ into close agreement with the reference $\textbf{X}^{\rm CCSD(T)}_{\rm VTZ}$.
Overall, while CCSD(T)/cc-pVDZ geometries may suffice for smaller systems, the geometry scheme offers an attractive and computationally efficient alternative for obtaining near-CCSD(T)/cc-pVTZ accuracy for electronically and structurally sensitive systems such as APs.

The accuracies of the L-CC2/aug-cc-pVDZ excitation energies obtained using geometries optimized with two DFT methods (B3LYP and $\omega$B97XD) and MP2, with the cc-pVTZ basis set, were also assessed using the same error metrics (MSE, MAE, and SDE). The L-CC2 energies calculated with $\textbf{X}^{\rm CCSD(T)}_{\rm VTZ}$ geometries from Table~\ref{tab:geom_benchmark} were taken as reference (see Table~\ref{tab:other_methods}). 
All three methods exhibit large errors relative to the reference data, with $\omega$B97XD geometries showing the best performance (MAE = 0.035~eV, SDE = 0.044~eV), followed by those of B3LYP (MAE=0.080~eV, SDE = 0.109~eV) and MP2 (MAE=0.095~eV, SDE = 0.138~eV). 
These results indicate that $\omega$B97XD provides the most reliable geometries among the three, and overall, both DFT methods outperform MP2. 
As reported in ~\RRef{majumdar2024influence}, DFT methods are capable of capturing pJT distortions, particularly in 4AP, whereas MP2 tends to favor the high-symmetry \dthreeh configuration. 

\begin{table*}[!htpb]
\centering
\caption{
L-CC2/aug-cc-pVDZ energies of the S$_1$ and T$_1$ states with respect to the S$_0$
ground state along with the singlet-triplet gap, S$_1$-T$_1$, are given in eV based on minimum energy
geometries obtained with DFT and MP2 methods. 
Molecular names are given along with point group symmetry in parentheses. 
Error metrics, in eV, compared to L-CC2/aug-pVDZ energies determined with CCSD(T)/cc-pVTZ (from Table~\ref{tab:geom_benchmark})
geometries are given in the convention "Method $-$ Reference": MSE: mean signed error,
MAE: mean absolute error, and
SDE: standard deviation of the error.
 }
\scriptsize
\addtolength{\tabcolsep}{1.2pt}
\begin{tabular}[t]{l rrrl rrrl rrr}
\hline
\multicolumn{1}{l}{System}   &
\multicolumn{4}{l}{$\textbf{X}^{\rm B3LYP}_{\rm VTZ}$ } & 
\multicolumn{4}{l}{$\textbf{X}^{\omega{\rm B97XD}}_{\rm VTZ}$ } & 
\multicolumn{3}{l}{$\textbf{X}^{\rm MP2}_{\rm VTZ}$ } \\
\cline{2-4} \cline{6-8}  \cline{10-12}
\multicolumn{1}{l}{} &
\multicolumn{1}{l}{S$_1$} &
\multicolumn{1}{l}{T$_1$} &
\multicolumn{1}{l}{S$_1$-T$_1$} &
\multicolumn{1}{l}{} &
\multicolumn{1}{l}{S$_1$} &
\multicolumn{1}{l}{T$_1$} &
\multicolumn{1}{l}{S$_1$-T$_1$} &
\multicolumn{1}{l}{} &
\multicolumn{1}{l}{S$_1$} &
\multicolumn{1}{l}{T$_1$} &
\multicolumn{1}{l}{S$_1$-T$_1$} \\
\hline 

cyclazine (\dthreeh) & $1.045$ & $1.177$ & $-0.132$ && $1.067$ & $1.195$ & $-0.128$ && $1.039$ & $1.169$ & $-0.130$ \\
cyclazine (\cthreeh) & $1.045$ & $1.177$ & $-0.132$ && $1.067$ & $1.195$ & $-0.128$ && $1.039$ & $1.169$ & $-0.130$ \\
2-aza (\ctwov) & $0.906$ & $0.992$ & $-0.086$ && $0.929$ & $1.011$ & $-0.082$ && $0.894$ & $0.976$ & $-0.082$ \\
2-aza (\cs) & $0.906$ & $0.992$ & $-0.086$ && $1.024$ & $1.051$ & $-0.027$ && $0.885$ & $0.965$ & $-0.080$\\
2,5-biaza (\ctwov) & $0.769$ & $0.840$ & $-0.071$ && $0.789$ & $0.856$ & $-0.067$ && $0.734$ & $0.803$ & $-0.069$\\
2,5-biaza (\cs) & $0.769$ & $0.840$ & $-0.071$ && $1.194$ & $1.037$ & $0.157$ && $0.734$ & $0.803$ & $-0.069$ \\
2,5,8-triaza (\dthreeh) & $0.621$ & $0.677$ & $-0.056$ && $0.652$ & $0.704$ & $-0.052$ && $0.580$ & $0.635$ & $-0.055$ \\
2,5,8-triaza (\cthreeh) & $0.875$ & $0.777$ & $0.098$ && $1.292$ & $1.007$ & $0.285$ && $0.580$ & $0.635$ & $-0.055$  \\
1,3,4,6-tetraaza (\ctwov) & $2.186$ & $2.324$ & $-0.138$ && $2.229$ & $2.354$ & $-0.125$ && $2.206$ & $2.336$ & $-0.130$ \\
1,3,4,6,7,9-hexaaza (\dthreeh) & $2.733$ & $2.956$ & $-0.223$ && $2.769$ & $2.987$ & $-0.218$ && $2.739$ & $2.964$ & $-0.225$ \\
MSE & $-0.132$ & $-0.054$ & $-0.077$ && $-0.016$ & $0.010$ & $-0.026$ && $-0.174$ & $-0.084$ & $-0.090$ \\
MAE & $0.154$ & $0.075$ & $0.080$ && $0.073$ & $0.038$ & $0.035$ && $0.185$ & $0.090$ & $0.095$ \\
SDE & $0.210$ & $0.102$ & $0.109$ && $0.083$ & $0.040$ & $0.044$ && $0.272$ & $0.134$ & $0.138$ \\
\hline 
\end{tabular}
\label{tab:other_methods}
\end{table*}%

\begin{figure}[!hbpt]
\centering
\includegraphics[width=\linewidth]{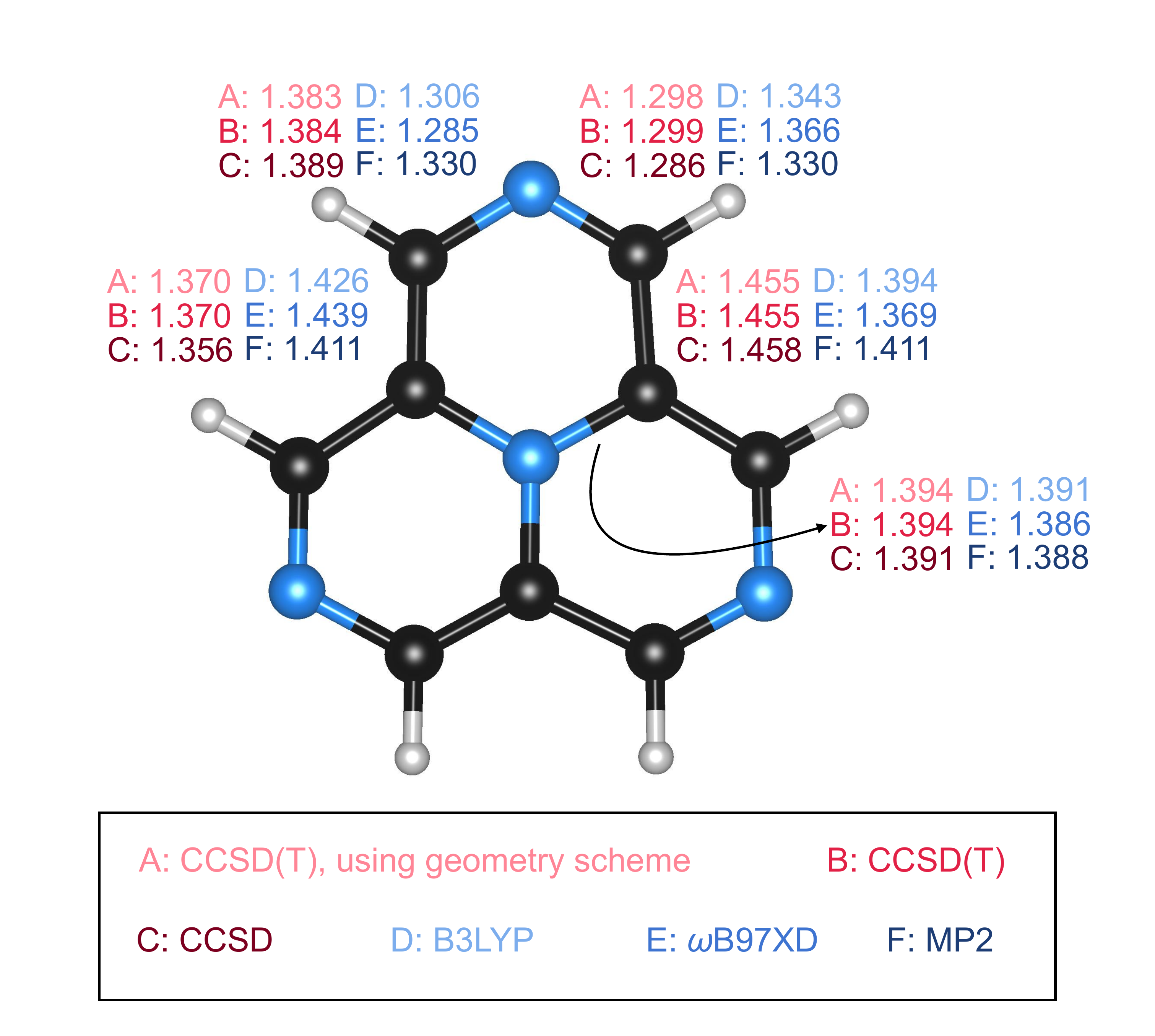}
\caption{
For 2,5,8-triazacyclazine (4AP) in the \cthreeh geometry, C–C and C–N bond lengths are shown for structures optimized using various methods with the cc-pVTZ basis set. All bond distances are given in~\AA. The MP2 results highlight its inability to capture the pseudo-Jahn--Teller distortion from \dthreeh to \cthreeh.
}
\label{4AP_C3h}
\end{figure}

Figure~\ref{4AP_C3h} compares the individual bond lengths of the \cthreeh configuration of 2,5,8-triazacyclazine (4AP) across different methods, showing that the MP2-optimized structure retains twofold rotational symmetry within the molecular plane, characteristic of the \dthreeh point group. In contrast, all other correlated methods, including the DFT methods B3LYP and $\omega$B97XD, correctly break this twofold symmetry in the low-symmetry \cthreeh configuration. 
The $\omega$B97XD functional also predicts symmetry lowering for 2-azacyclazine (2AP) and  2,5-biazacyclazine (3AP), while in the case of cyclazine (1AP) such distortions are reproduced only by post-MP2 methods (CCSD and CCSD(T); see Table~\ref{tab:geom_benchmark}). Therefore, for systems such as 2AP, 3AP, and 4AP that exhibit topologically destabilizing effects, $\omega$B97XD is preferable to MP2 for geometry optimization.

The remainder of this study employs the extrapolated geometries, $\textbf{X}^{\rm CCSD(T),\,\Delta_{DT}}_{\rm VTZ}$, to compute STGs at the L-CC2/aug-cc-pVDZ level across the chemical space of 104 APs and to investigate pJT symmetry lowering in systems that have not been previously examined.

\subsection{Pseudo-Jahn--Teller (PJT) distortion and the impact on singlet-triplet gap of azaphenalenes}\label{sec:pJT}

We examined pJT distortions across the 104 APs using the basis set extrapolation scheme for geometries as described in Section~\ref{sec:geom_scheme} to obtain equilibrium geometries at the CCSD(T)/cc-pVTZ level, $\textbf{X}^{\rm CCSD(T),\,\Delta_{DT}}_{\rm VTZ}$, constrained to their maximal symmetry point groups (\dthreeh, \cthreeh, \ctwov, or \cs). The selected Z-matrix coordinates allow in-plane distortions leading to \dthreeh~$\rightarrow$~\cthreeh and \ctwov~$\rightarrow$~\cs symmetry lowering; further in-plane distortions to the subgroups of \cthreeh~or~\cs were not observed. While out-of-plane distortions in the excited state (S$_1$) have been reported to play a role in the INVEST character~\cite{karak2024reverse}, the present work focuses on in-plane (soft-mode) distortions of the ground state (S$_0$) of APs.

Following ~\RRef{majumdar2024influence}, constrained MP2/cc-pVDZ optimizations were performed by varying two symmetry-defining bond lengths along the \dthreeh~$\rightarrow$~\cthreeh and \ctwov~$\rightarrow$~\cs paths. Single-point CCSD(T)/cc-pVDZ energies were then computed along both the $r_1 = r_2$ line (high-symmetry) and $r_1 \neq r_2$ region (low-symmetry) to locate minima. Low-symmetry minima correspond to true pJT-distorted geometries.

\begin{table}[!hbtp]
\caption{
For thirteen azaphenalenes undergoing pseudo-Jahn--Teller distortion and symmetry lowering, singlet–triplet gaps (S$_1$–T$_1$) are shown as predicted by the L-CC2 method with the aug-cc-pVDZ basis set. Values corresponding to the low-symmetry geometries are given in parentheses. 
The automerization barrier, $E^\ddagger$, corresponds to the high-symmetry saddle point (maximum) along the reaction pathway connecting two equivalent low-symmetry minima. This quantity was determined using two-point CBS extrapolation of CCSD(T) energies with the cc-pVTZ and cc-pVQZ basis sets, and is reported in kJ~mol$^{-1}$.
Harmonic vibrational frequencies calculated for the high-symmetry configuration at the $\omega$B97XD/cc-pVTZ level are also listed; except for cyclazine (1AP), a single imaginary frequency corresponding to the unstable normal mode was noted.
}
\footnotesize
 \begin{tabular}[t]{lll lll}
\hline
\multicolumn{1}{l}{Name} & \multicolumn{2}{l}{ S$_1$-T$_1$} &  \multicolumn{2}{l}{$E^\ddagger$} & \multicolumn{1}{l}{$\overline{\nu}_{\rm imag}$} \\
\hline
cyclazine$^a$   & $-0.134$ ($-0.047$) && 0.3 && 159$^c$\\
2-aza$^b$   &  $-0.089$ ($+0.095$) && 2.3&& 533$i$\\
2,5-biaza$^b$    & $-0.071$ ($+0.224$) && 5.5&& 907$i$\\
2,5,8-triaza$^a$    & $-0.058$ ($+0.337$) && 11.1&& 1236$i$\\ 
1,2,5,6-tetraaza$^b$ & $-0.010$ ($+0.126$) &&  1.0&& 424$i$\\ 
1,2,8,9-tetraaza$^b$ &  $-0.089$ ($+0.009$)  & & 0.0  && 177$i$ \\
1,2,3,5,8-pentaaza$^b$  & $-0.066$ ($+0.147$) && 2.5&& 633$i$\\
1,2,5,6,8-pentaaza$^b$ & $-0.047$ ($+0.188$) && 3.4&& 710$i$\\ 
1,2,5,8,9-pentaaza$^b$  &  $-0.022$ ($+0.188$) && 2.3&& 631$i$\\
1,2,3,4,5,6,8-heptaaza$^b$   & $-0.097$ ($+0.050$) && 0.6&& 424$i$\\ 
1,2,3,4,5,8,9-heptaaza$^b$ &  $-0.045$ ($+0.116$) && 0.7&& 403$i$\\
1,2,3,5,6,7,8-heptaaza$^b$ &  $-0.099$ ($+0.083$) && 1.5&& 467$i$\\ 
1,2,3,4,5,6,7,8,9-nonaaza$^a$  & $-0.188$ ($+0.296$) && 1.8&& 444$i$\\
\hline

\hline
\end{tabular}
\begin{tablenotes}
 \item 
 \footnotesize{$^a$ \dthreeh $\rightarrow$ \cthreeh},$\quad$  
 \footnotesize{$^b$  \ctwov $\rightarrow$  \cs } \\ 
 \footnotesize{$^c$ $\omega$B97XD predicts \dthreeh configuration as a minimum}
 \end{tablenotes}
\label{tab:dist_gaps}
\end{table}

Systems were classified as pJT-active if two criteria were met:  
(1) $|r_1 - r_2| \geq 0.02$~\AA~at the asymmetric minimum, and 
(2) the energy difference ($\Delta E$) between the lowest energy point in the $r_1=r_2$ line and
the lowest energy point of the $r_1 \neq r_2$ region, satisfies $\Delta E \geq 0.1$~kJ~mol$^{-1}$.
Harmonic frequency analyses at the $\omega$B97XD/cc-pVTZ level were also used to identify borderline cases. One such system, 1,2,8,9-tetraazacyclazine, showed $|r_1 - r_2| < 0.02$~\AA, $\Delta E < 0.1$~kJ~mol$^{-1}$ but exhibited a single imaginary in-plane mode (177.3$i$ cm$^{-1}$), confirming its structural instability in its \ctwov geometry.

For the 13 systems that met the pJT criteria (Table~\ref{tab:dist_gaps}), their final low-symmetry geometries were also obtained using the geometry scheme. Figure~\ref{fig:sites} shows the distribution of all APs by point-group symmetry and N substitution pattern in the periphery, with arrows marking cases of symmetry lowering. Cyclazine, which is also weakly pJT-active, undergoing a \dthreeh$\rightarrow$\cthreeh distortion, is excluded since it lacks peripheral N atoms.

Table~\ref{tab:dist_gaps} also reports automerization barriers ($E^\ddagger$), defined as the energy of the high-symmetry saddle point connecting two equivalent low-symmetry minima. These barriers were obtained via two-point CBS extrapolation of CCSD(T) energies with cc-pVTZ and cc-pVQZ basis sets calculated on $\textbf{X}^{\rm CCSD(T),\,\Delta_{DT}}_{\rm VTZ}$. All systems except cyclazine show one imaginary frequency (at the $\omega$B97XD/cc-pVTZ level using $\textbf{X}^{\rm CCSD(T),\,\Delta_{DT}}_{\rm VTZ}$) at the high-symmetry structure, corresponding to the unstable in-plane mode. Cyclazine instead exhibits a small real frequency (159~cm$^{-1}$), reflecting the weak pJT coupling not captured at the DFT-level.

The large magnitudes of imaginary frequencies for the other systems highlight strong vibronic coupling and the need for careful geometry verification before interpreting INVEST character, particularly when using MP2 or other approximate methods for their geometry optimization. For all systems in Table~\ref{tab:dist_gaps} except cyclazine, the L-CC2/aug-cc-pVDZ STGs change from inverted (negative STG) to non-inverted (positive STG) upon relaxation. 
This behavior is observed for all three electron-deficient systems
(2-azacyclazine, 2AP; 2,5-biazacyclazine, 3AP; 2,5,8-triazacyclazine, 4AP) as well
as for the nine mixed-substitution systems, demonstrating that the STG
inversions present at the highest-symmetry geometries are removed by symmetry-lowering distortions.

The high-symmetry geometries of all 104 APs and the low-symmetry geometries of thirteen distortion-prone APs are provided for benchmarking and reproducibility (see Data Availability).

\begin{figure*}[!htpb]
\centering
\includegraphics[width=\linewidth]{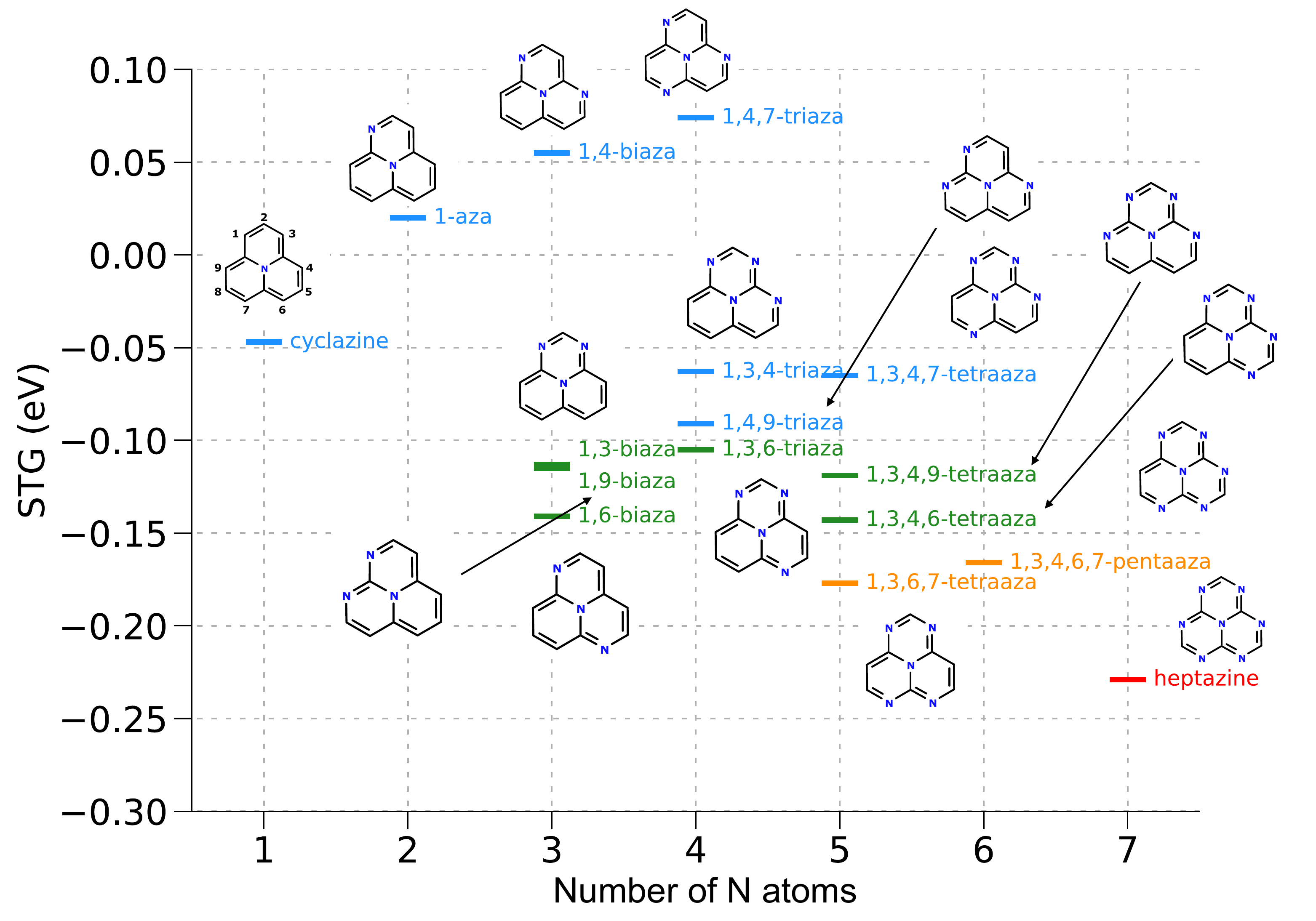}
\caption{
L-CC2/aug-cc-pVDZ singlet-triplet gaps of cyclazine (1AP) and its fifteen topologically charge-stabilized polyaza systems with N substituting the electron-rich carbon centers. 1,3,4,6-tetraazacyclazine (5AP) is the only unsubstituted azaphenalene for which the singlet-triplet has been experimentally measured (0-0 value is $-$0.047 eV)\cite{wilson2024spectroscopic}. Heptazine (7AP) corresponds to 1,3,4,6,7,9-hexaazacyclazine.  Names of molecules with singlet-triplet gaps in the range $<-0.2$ eV, $\left(-0.2 , -0.15\right]$ eV,  $\left(-0.15 , -0.1\right]$ eV, and $> -0.1$ eV are shown in red, orange, green, and blue. 
}
\label{trends}
\end{figure*}

\begin{figure*}[!hbtp]
\centering
\includegraphics[width=0.8\linewidth]{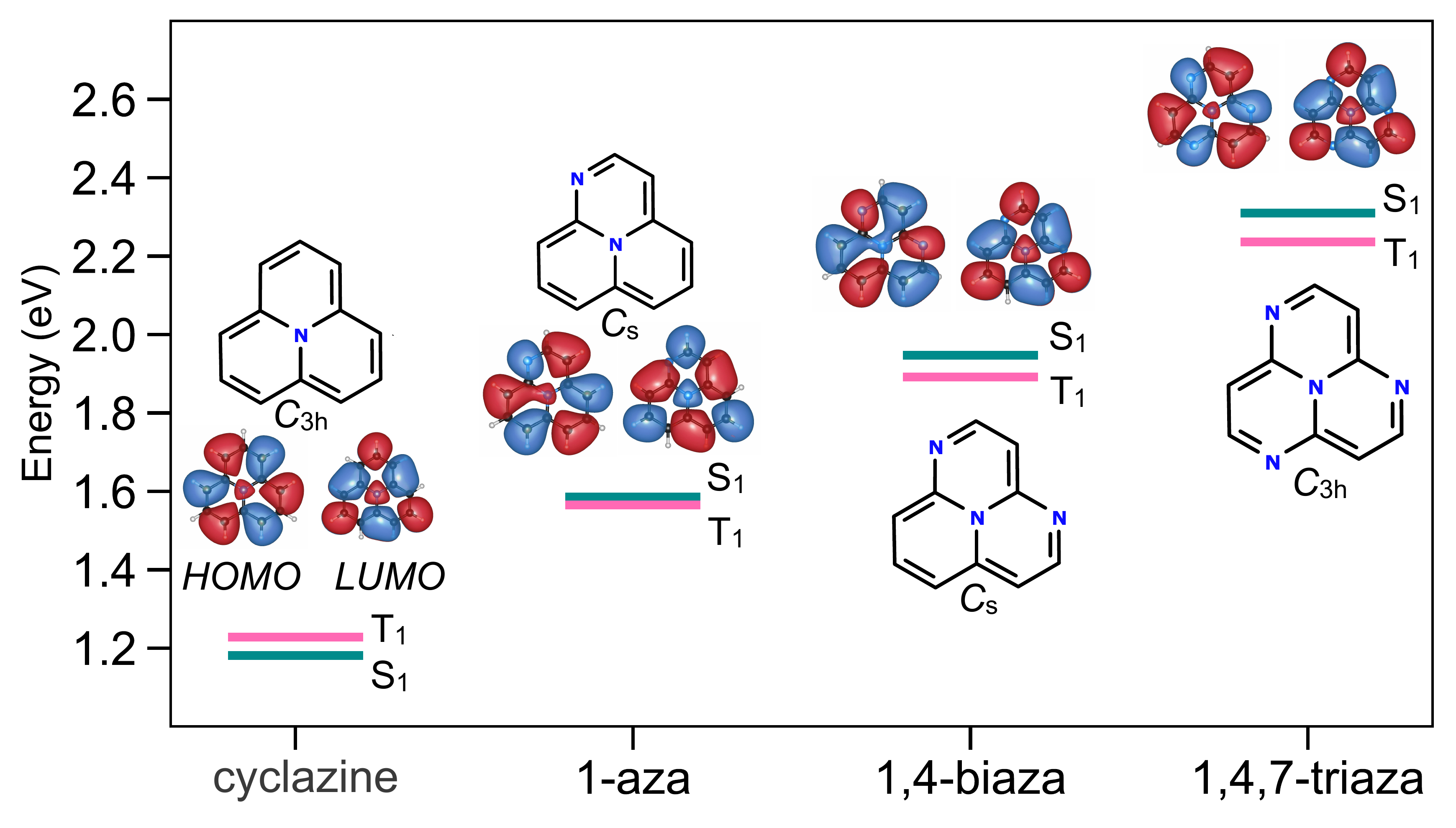}
\caption{
Excitation energies of S$_1$ and T$_1$ states relative to the S$_0$ energy (in eV) and
 frontier molecular orbitals of cyclazine (1AP), 1-azacyclazine, 1,4-biazacyclazine and 1,4,7-triazacyclazine calculated with $\omega$B97XD/cc-pVTZ are plotted with an isovalue of 0.01 a.u.. In all cases, we used the 
extrapolated geometries, $\textbf{X}^{\rm CCSD(T),\,\Delta_{DT}}_{\rm VTZ}$.
White$|$black$|$blue atoms denote H$|$C$|$N.
}
\label{MO}
 \end{figure*}

\subsection{Effect of symmetry and substitution pattern on singlet-triplet gaps\label{sec:effectofsymm}}

As discussed in Section~\ref{sec:survey}, among the 15 APs featuring N substitution at topologically charge-stabilizing (electron-rich or HOMO-dominated) sites, 14 have been investigated previously, with the exception of 1,4,9-triazacyclazine. By contrast, only three systems (2AP, 3AP, and 4AP) contain N atoms at electron-deficient (LUMO) sites, and none of these exhibit negative STGs at their low-symmetry minima following pJT distortion. The remaining 85 APs feature mixed substitution patterns that are less straightforward to generalize. Hence, we focus on cyclazine and 15 APs with N substitution at electron-rich (HOMO) sites of cyclazine's periphery, which offers a chemically intuitive and structurally consistent platform for identifying trends and design principles underlying Hund’s rule violation in the AP family.

Figure~\ref{trends} presents the STGs of these topologically charge-stabilized APs as a function of the number of N atoms. Although no simple linear correlation is observed across the 16 systems (including cyclazine), several systematic trends emerge that provide insight into structure–property relationships relevant for molecular design. A clear trend is observed along the series cyclazine (1AP) $\rightarrow$ 1,6-biazacyclazine $\rightarrow$ 1,3,6,7-tetraazacyclazine $\rightarrow$ heptazine (7AP), corresponding to a progressive decrease in the STG. These systems define the lower bound of STG values for a given number of N substitutions, with 7AP exhibiting the most negative STG.

The 1,3,4,6-tetraazacyclazine isomer (5AP), previously studied by Wilson et al.~\cite{wilson2024spectroscopic}, provides direct experimental evidence for an inverted singlet–triplet ordering. Interestingly, two other constitutional isomers of 5AP, namely, 1,3,6,7-tetraazacyclazine and 1,3,4,9-tetraazacyclazine, exhibit distinct magnitudes of inversion despite sharing the same \ctwov symmetry. Among them, 1,3,6,7-tetraazacyclazine shows a more pronounced negative STG than 5AP, whereas 1,3,4,9-tetraazacyclazine displays a smaller inversion, highlighting the sensitivity of the STG to substitution-induced perturbations in the electronic structure.

Another interesting branch starts from cyclazine and proceeds through 1-azacyclazine, 1,4-biazacyclazine, and 1,4,7-triazacyclazine, showing the opposite trend, i.e., a gradual increase in the STG. The underlying reason is illustrated by the frontier MOs in Figure~\ref{MO}. In 1-aza and 1,4-biaza, the HOMO exhibits progressively greater electron density at site-2, which corresponds to a LUMO site in the parent compound, cyclazine. This redistribution enhances the spatial overlap between the HOMO and LUMO, increasing their exchange coupling. In 1,4,7-triazacyclazine, N substitution at positions 2, 5, and 8 further amplifies the HOMO density at these sites, resulting in the greatest HOMO–LUMO overlap within this subset. The consequent increase in exchange interaction stabilizes the T$_1$ state relative to S$_1$, thereby producing larger (positive) STGs.

\subsection{Hund's rule violating azaphenalenes \label{sec:topsystems}}

Among the 104 APs investigated, 13 exhibit pJT symmetry lowering, resulting in a total of 117 distinct configurations (for the 13 systems both high and low symmetry structures are included).  
Singlet and triplet excitation energies and corresponding STGs for the 30 APs exhibiting negative STGs, computed at the L-CC2/aug-cc-pVDZ level using the extrapolated equilibrium geometries, $\textbf{X}^{\rm CCSD(T),\,\Delta_{DT}}_{\rm VTZ}$, are listed in Table~\ref{tab:top30_negative}. 

Of the 30 INVEST APs listed in Table~\ref{tab:top30_negative}, 1,3,4,6,7,9-hexaazacyclazine (heptazine, 7AP) with \dthreeh symmetry displays the most negative STG ($-0.229$~eV). The remaining systems comprise 15 molecules of \ctwov symmetry, 13 of \cs symmetry, along with cyclazine (\cthreeh). The occurrence of negative STGs in \cs structures indicates that high molecular symmetry is not a prerequisite for achieving Hund’s rule violation. 
The oscillator strengths of the first excited singlet state for the 30 INVEST
systems are also listed in Table~\ref{tab:top30_negative}.
For nearly all molecules, the oscillator strengths are negligible; the only
exception is 1,2,3,4,6,7-hexaazacyclazine, which exhibits a small but non-negligible value of 0.01 a.u.

\begin{table}[h]
\caption{
The energies of S$_1$ and T$_1$ states along with singlet-triplet gaps (S$_1$$-$T$_1$) in eV for the thirty azaphenalenes calculated using L-CC2/aug-cc-pVDZ on the CCSD(T)/cc-pVTZ-quality  
minimum energy geometries determined using the geometry scheme, $\textbf{X}^{\rm CCSD(T),\,\Delta_{DT}}_{\rm VTZ}$. 
The ${\rm S}_0 \rightarrow {\rm S}_1$ oscillator strengths ($f$, in atomic units) are given in
parentheses next to the ${\rm S}_1$ energies.
The molecules are arranged in the order of most to least negative STG. 
}
\footnotesize  
 \begin{tabular}[t]{l l l l}
\hline
 Molecule (Symmetry)                       &  S$_1$ (\textit{f})    &  T$_1$    &  S$_1$$-$T$_1$        \\
\hline
1,3,4,6,7,9-hexaaza  (\dthreeh)     &  $2.723$ (0.000)  &  $2.952$  &  $-0.229$  \\
1,3,6,7-tetraaza     (\ctwov)     &  $2.127$ (0.000) &  $2.304$  &  $-0.177$  \\
1,3,4,6,7-pentaaza     (\cs)      &  $2.466$ (0.002) &  $2.632$  &  $-0.166$  \\
1,6-biaza             (\ctwov)     &  $1.560$ (0.001)  &  $1.701$  &  $-0.141$  \\
1,3,4,6-tetraaza     (\ctwov)           &  $2.190$ (0.002) &  $2.326$  &  $-0.136$  \\
1,2,3,4,9-pentaaza (\ctwov)       &  $1.914$ (0.000) &  $2.049$  &  $-0.135$  \\
1,2,4,6,8,9-hexaaza (\ctwov)       &  $1.779$ (0.001) &  $1.901$  &  $-0.122$  \\
1,3,4,9-tetraaza (\ctwov)       &  $2.126$ (0.006) &  $2.245$  &  $-0.119$  \\
1,6,8-triaza (\ctwov)          &  $1.434$ (0.000)  &  $1.551$  &  $-0.117$  \\
1,3-biaza (\ctwov)         &  $1.643$ (0.003) &  $1.758$  &  $-0.115$  \\
1,9-biaza (\ctwov)          &  $1.583$ (0.003)  &  $1.696$  &  $-0.113$  \\
1,2,3,4,5,6-hexaaza (\ctwov)          &  $1.722$ (0.001) &  $1.833$  &  $-0.111$  \\
1,3,6-triaza        (\cs)      &  $1.968$ (0.001) &  $2.073$  &  $-0.105$  \\
1,2,3-triaza (\ctwov)          &  $1.412$ (0.000) &  $1.510$  &  $-0.098$  \\
1,2,3,6,7-pentaaza (\ctwov)         &  $1.831$ (0.006) &  $1.926$  &  $-0.095$  \\
1,4,9-triaza       (\cs)   &  $1.924$ (0.003) &  $2.015$  &  $-0.091$  \\
1,2,4,6,9-pentaaza     (\cs)   &  $2.016$ (0.002) &  $2.095$  &  $-0.079$  \\
1,2,3,4,6,7-hexaaza    (\cs)      &  $2.200$ (0.010) &  $2.278$  &  $-0.078$  \\
1,2,3,4,6,9-hexaaza    (\cs)     &  $2.323$ (0.001) &  $2.400$  &  $-0.077$  \\
1,3,4,7-tetraaza    (\cs)     &  $2.338$ (0.003) &  $2.403$  &  $-0.065$  \\
1,2,3,4,6-pentaaza    (\cs)       &  $1.979$ (0.002) &  $2.042$  &  $-0.063$  \\
1,3,4-triaza    (\cs)       &  $1.986$ (0.005) &  $2.049$  &  $-0.063$  \\
1,2,6,9-tetraaza    (\cs)        &  $1.655$ (0.005) &  $1.709$  &  $-0.054$  \\
cyclazine                  (\cthreeh)    &  $1.181$ (0.000) &  $1.228$  &  $-0.047$  \\
1,3,4,6,8-pentaaza    (\ctwov)  &  $2.045$ (0.007)  &  $2.091$  &  $-0.046$  \\
1,5,9-triaza         (\ctwov)     &  $1.415$ (0.007)  &  $1.444$  &  $-0.029$  \\
1,3,5-triaza        (\cs)     &  $1.560$ (0.004) &  $1.587$  &  $-0.027$  \\
1,2,3,4-tetraaza    (\cs)    &  $1.825$ (0.004) &  $1.851$  &  $-0.026$  \\
1,3,5,8-tetraaza   (\ctwov)    &  $1.359$ (0.005) &  $1.382$  &  $-0.023$  \\
1,3,5,7-tetraaza    (\cs)   &  $1.932$ (0.002) &  $1.940$  &  $-0.008$  \\
\hline
\end{tabular}
\label{tab:top30_negative}
\end{table}

Notably, cyclazine (1AP) is the only system that undergoes pJT distortion while retaining a negative STG, as summarized in Table~\ref{tab:dist_gaps}. 
As stated before, none of the three APs with N atoms exclusively at electron-deficient 
(LUMO) sites (2AP, 3AP, and 4AP) show inverted gaps, and thus they are absent from Table~\ref{tab:top30_negative}. 
Among the 15 APs with N atoms exclusively at electron-rich (HOMO) sites (see Figure~\ref{trends}), 12 exhibit negative STGs. Of these, 7AP has the most negative STG of $-0.229$~eV, while 1,3,4-triazacyclazine (\cs) has the least negative STG of $-0.063$ eV.

Further, of the 30 INVEST-type APs, 17 originate from the subset of 85 systems featuring mixed substitution patterns, where N atoms occupy both electron-rich (HOMO) and electron-deficient (LUMO) sites of the parent compound cyclazine (see Figure~\ref{fig:sites}). All 17 exhibit negative STGs at their equilibrium geometries, corresponding to high-symmetry structures permitted by their stoichiometry, i.e., none of these systems undergo pJT distortion. These 17 species can be regarded as derivatives of the topologically charge-stabilizing systems that exhibit negative STGs in Figure~\ref{trends}, and their trends can be rationalized by comparison with their respective parent compounds. For consistency, the IUPAC convention of assigning the smallest possible indices to the heteroatoms has been followed when identifying parent–derivative relationships; for example, 1,3 or 1,3,6 are preferred over 1,4 or 1,4,6, respectively (see cyclazine in Figure~\ref{trends} for the numbering scheme).

When comparing the S$_1$ and T$_1$ excitation energies, all 17 derivatives show less negative values than their parents, indicating that substitution of N at a LUMO site of cyclazine narrows the HOMO–LUMO gap. In the extreme cases, such as 2-azacyclazine (2AP), 2,5-biazacyclazine (3AP), and  2,5,8-triazacyclazine (4AP), this narrowing can enhance pJT coupling, as discussed in ~\RRef{majumdar2024influence}. Among the 17 mixed systems, only 1,2,3,4,9-pentaazacyclazine (\ctwov) exhibits a more negative STG ($-0.135$~eV) than its parent 1,3,4,9-tetraazacyclazine (\ctwov, STG = $-0.119$~eV), while the remaining 16 systems show larger (less negative) gaps than their respective parents. However, this apparent deviation is small compared to the uncertainty of the L-CC2 method (MAE = 0.016~eV, SDE = 0.013~eV for systems of similar size and topology, see Section~\ref{sec:comp_details} for more details). Hence, both systems can be considered to possess effectively similar STGs within the predictive uncertainty of the method.

\section{Conclusions}

The fifth-generation OLEDs rely critically on the inverted singlet–triplet gaps (STGs)\cite{tuvckova2022origin}, a unique feature of light-emitting molecules that operate via delayed fluorescence from inverted singlet and triplet states (DFIST)~\cite{aizawa2022delayed}. Molecules exhibiting this behavior are termed inverted singlet-triplet energy gap (INVEST) systems\cite{li2022organic,loos2023heptazine,loos2025correction}. Achieving INVEST depends on a delicate interplay of molecular topology, electronic structure, and geometric stability. Despite intense research interest, only a handful of molecules have been experimentally confirmed to possess negative STGs, among which azaphenalenes (APs) represent the most promising and systematically studied class. In this work, we present a comprehensive enumeration and analysis of all possible N-substituted APs, reporting 104 unique molecules. These were classified as topologically charge-stabilizing (N substitution at electron-rich sites), topologically charge-destabilizing (N substitution at electron-deficient sites), or mixed, depending on the location of the N atoms in the periphery. We also compiled prior studies on this molecular family, noting that only 24 of the 104 possible APs have been explored before.

We highlighted the critical role of geometry optimization in accurately determining STGs and, hence, in identifying genuine INVEST systems. 
While pseudo–Jahn--Teller (pJT) distortions in the closed-shell electronic ground state already require careful treatment, the physically relevant quantities for INVEST are the adiabatic (0–0) STGs, which are particularly sensitive to symmetry-lowering distortions in the open-shell excited states. 
Such pJT distortions in excited states are generally more prevalent and pronounced than in the ground state and can qualitatively alter the STG upon relaxation. Vertical STGs evaluated at high-symmetry geometries therefore provide only an initial indication of a possible inversion, whereas reliable conclusions require adiabatic gaps obtained at fully relaxed geometries.
The pJT distortions inherent in several APs are not captured by methods such as B3LYP and MP2\cite{majumdar2024influence}. For distortion-prone systems, only CCSD(T)-quality geometries provide reliable, stable structures. To achieve this accuracy efficiently, we employed a geometry extrapolation scheme based on basis-set additivity, enabling near-CCSD(T)/cc-pVTZ precision at significantly reduced computational cost. 
Among the 104 APs, thirteen undergo symmetry lowering from their high-symmetry configurations to low-symmetry minima, resulting in a total of 117 distinct geometries analyzed.

Cyclazine, the parent member with a single central N, serves as a special case. It is the only system that retains a negative STG, even after pJT distortion to its low-symmetry minimum (\cthreeh). All other APs possess peripheral N atoms, displaying diverse electronic effects. Among the 15 systems with N substitution exclusively at electron-rich centers, 12 exhibit negative STGs at their equilibrium geometries. Of these, 1,4,9-triazacyclazine emerges as a previously unexplored member with potential INVEST character. Analysis across the 15 topologically stabilized systems further reveals that high-symmetry members (\dthreeh~and~\ctwov) display minimal HOMO–LUMO overlap, akin to cyclazine, while some of the lower-symmetry species (\cthreeh~and~\cs) show greater frontier MO overlap, resulting in larger, positive STGs.
All three APs with N atoms solely at electron-deficient sites undergo pJT distortion, losing their inverted gap behavior. 
The remaining 85 systems feature mixed substitution patterns, of which only six have been explored before.

Excitation energies and STGs of all 117 systems, derived from 104 unique APs, were evaluated using the L-CC2/aug-cc-pVDZ method, which provides benchmark-level accuracy for INVEST molecules at moderate computational expense \cite{majumdar2025leveraging}. Among the 13 distortion-prone systems, only cyclazine maintains a negative vertical STG, while all others yield positive values in their minima. Across the entire set, heptazine (7AP) displays the most negative STG ($-0.229$~eV), consistent with prior reports \cite{loos2023heptazine,ehrmaier2019singlet}. Interestingly, 1,3,6,7-tetraazacyclazine (STG=$-0.177$~eV) exhibits a more negative gap than its constitutional isomer 1,3,4,6-tetraazacyclazine (5AP, STG=$-0.163$~eV) that has been experimentally verified\cite{wilson2024spectroscopic}. 1,3,4,6,7-pentaazacyclazine (6AP, \cs) and 1,6-biazacyclazine (\ctwov) also feature STGs less negative than 7AP but more negative than 5AP.

Overall, 30 APs are identified here as potential INVEST systems, providing concrete molecular cores for the design of next-generation OLED emitters with inverted singlet–triplet ordering. Of these, only a few have been previously studied. Future extensions of this work should include 0–0 corrections to refine predicted gaps and a detailed examination of excited-state dynamics. The choice of geometry optimization method remains critical, especially for systems exhibiting pJT activity. 
Functionalized APs such as those of 5AP~\cite{kusakabe2024inverted} and 7AP~\cite{aizawa2022delayed,actis2023singlet} offer natural starting points for experimental exploration. While this study focuses on N-substitution, the observed structure–property trends can be generalized to other heteroatom substitutions or other functionalizations. Realizing such broader chemical spaces will require integration of evolutionary design strategies with high-throughput quantum chemistry.

\section{Supplementary Information}
i) Geometries and additional data are available in the AP117 dataset~\cite{AP117},  
ii) Table S1 presents names and SMILES representations, 
iii) Table S2 presents excitation energies and singlet–triplet gaps.

\section{Data Availability}
The data that support the findings of this study are
within the article and its supplementary material.

\section{Acknowledgments}
We acknowledge the support of the 
Department of Atomic Energy, Government
of India, under Project Identification No.~RTI~4007. 
All calculations have been performed using the Helios computer cluster, 
which is an integral part of the MolDis 
Big Data facility, 
TIFR Hyderabad \href{http://moldis.tifrh.res.in}{(http://moldis.tifrh.res.in)}.

\section{Author Declarations}

\subsection{Author contributions}
\noindent 
{\bf AM}: Conceptualization (equal); 
Analysis (equal); 
Data collection (equal); 
Writing (equal).
{\bf RR}: Conceptualization (equal); 
Analysis (equal); 
Data collection (equal); 
Funding acquisition; 
Project administration and supervision; 
Resources; 
Writing (equal).

\subsection{Conflicts of Interest}
The authors have no conflicts of interest to disclose.

\section{References}
\bibliography{ref} 
\end{document}